\documentclass[aps,showpacs,preprintnumbers,amsmath, amssymb]{revtex4}

\usepackage{float}
\usepackage{graphics,epsfig}
\usepackage{graphicx}
\usepackage{dcolumn}
\usepackage{bm}

\begin{document}
\baselineskip=0.8 cm
\title{{\bf Revisiting holographic superconductors with hyperscaling violation}}

\author{Qiyuan Pan$^{1,2,3}$\footnote{panqiyuan@126.com} and Shao-Jun Zhang$^{1}$\footnote{sjzhang84@hotmail.com}}
\affiliation{$^{1}$Instituto de F\'{\i}sica, Universidade de S\~{a}o
Paulo, C.P. 66318, 05315-970, S\~{a}o Paulo, Brazil}
\affiliation{$^{2}$Department of Physics, Key Laboratory of Low
Dimensional Quantum Structures and Quantum Control of Ministry of
Education, Hunan Normal University, Changsha, Hunan 410081, China}
\affiliation{$^{3}$State Key Laboratory of Theoretical Physics,
Institute of Theoretical Physics, Chinese Academy of Sciences,
Beijing 100190, China}

\vspace*{0.2cm}
\begin{abstract}
\baselineskip=0.6 cm
\begin{center}
{\bf Abstract}
\end{center}

We investigate the effect of the hyperscaling violation on the
holographic superconductors. In the s-wave model, we find that the
critical temperature decreases first and then increases as the
hyperscaling violation increases, and the mass of the scalar field
will not modify the value of the hyperscaling violation which gives
the minimum critical temperature. We analytically confirm the
numerical results by using the Sturm-Liouville method with the
higher order trial function and improve the previous findings in J.
High Energy Phys. {\bf 09}, 048 (2013). However, different from the
s-wave case, we note that the critical temperature decreases with
the increase of the hyperscaling violation in the p-wave model. In
addition, we observe that the hyperscaling violation affects the
conductivity of the holographic superconductors and changes the
expected relation in the gap frequency in both s-wave and p-wave
models.

\end{abstract}

%\keywords{AdS/CFT correspondence, Holographic superconductors, Hyperscaling violation}

\maketitle
\newpage
\vspace*{0.2cm}

\section{Introduction}

The anti-de Sitter/conformal field theories (AdS/CFT) correspondence
\cite{Maldacena,Witten,Gubser1998}, which relates a $d$-dimensional
quantum field theory with its dual gravitational theory that lives
in $(d+1)$ dimensions, has been employed to gain a better
understanding of the high $T_{c}$ superconductor systems from the
gravitational dual, for reviews, see Refs.
\cite{HartnollRev,HerzogRev,HorowitzRev,CaiRev} and references
therein. It was suggested that the instability of the bulk black
hole corresponds to a second order phase transition from normal
state to superconducting state which brings the spontaneous $U(1)$
symmetry breaking \cite{GubserPRD78}, and the properties of a
($2+1$)-dimensional superconductor can indeed be reproduced in the
($3+1$)-dimensional holographic dual model where the AdS black hole
geometry corresponds to a relativistic CFT at finite temperature
\cite{HartnollPRLJHEP}. Since many condensed matter systems do not
have relativistic symmetry, it is of great interest to construct the
corresponding holographic superconductor models by using the
nonrelativistic version of the AdS/CFT correspondence. The
holographic superconductors in the Lifshitz black hole spacetime
were constructed in Refs. \cite{EUTT,SinXuZhou} and further
investigated in Ref. \cite{BuPRD}. It was observed that the Lifshitz
black hole geometry results in different asymptotic behaviors of
temporal and spatial components of gauge fields than those in the
Schwarzschild-AdS black hole, which brings some new features of
holographic superconductor models \cite{BuPRD}. Considering the
holographic superconductors with Ho\v{r}ava-Lifshitz black holes, it
was found that the holographic superconductivity is a robust
phenomenon associated with asymptotic AdS black holes and the ratio
of the gap frequency to the critical temperature is a little larger
than the one in the relativistic situations \cite{CaiZhangHL2010}.
Holographic superconductor models in the Lifshitz-like geometry can
also be found, for example, in Refs.
\cite{Schaposnik,Abdalla,Tallarita,LalaPLB2014,ZPJPLB2014,LuWuNPB,GuoShu,
MomeniLifshitz,KEA2015,DectorNPB,JingPLB2015}.

Recently, the so-called hyperscaling violation metric
\cite{CharmousisJHEP,GouterauxJHEP,DHKTW,ACYJHEP,Ganjali} which can
be considered as an extension of the Lifshitz metric has received
considerable attention due to its potential applications to the
condensed matter physics
\cite{HuijseJHEP,FanJHEPPRD,LucasPRD,ChenGuoShu,ZhangPanAbdalla,KuangWu,DRoychowdhury}.
Besides the anisotropic (Lifshitz) scaling characterized by the
dynamic critical exponent $z>1$ \cite{KachruPRD}, the hyperscaling
violation can also be used to describe holographically the realistic
condensed matter systems with scaling properties going beyond the
standard Lorentz scaling at criticality \cite{KuangJHEP}. So it
seems to be an interesting study to explore the effect of the
hyperscaling violation on the holographic superconductors. More
recently, the author of Ref. \cite{FanJHEP} introduced the
holographic s-wave superconductors with hyperscaling violation and
numerically found that the superconductivity still exists for the
hyperscaling violation exponent $\theta=1$. Using the analytical
Sturm-Liouville (S-L) method, which was first proposed in
\cite{Siopsis} and later generalized to study holographic
insulator/superconductor phase transition in \cite{CaiSL}, it was
found that the critical temperature increases as the hyperscaling
violation increases \cite{FanJHEP}. However, using the shooting
method \cite{HartnollPRLJHEP} to numerically study these holographic
dual models in the full parameter space, we find that with the
increase of the hyperscaling violation, the critical temperature
decreases first and then increases. Obviously, the analytical result
given in Ref. \cite{FanJHEP} misses some important information and
does not agree well with our numerical calculation. Thus, we will
reinvestigate the holographic s-wave superconductors with
hyperscaling violation in this work. In addition to giving a more
complete picture of how the hyperscaling violation affects the
condensation for the scalar operator and the conductivity, we will
present an interesting fact that the value of the hyperscaling
violation, which gives the minimum critical temperature, remains
unchanged even we vary the mass of the scalar field. Furthermore,
comparing with the second order trial function used in Ref.
\cite{FanJHEP}, we will improve the S-L method by including the
higher order terms in the expansion of the trial function to reduce
the disparity between the analytical and numerical results, and
what's more, to obtain the analytical results which are completely
consistent with the numerical findings.

Besides the holographic s-wave superconductor model, a holographic
p-wave model can be realized by introducing a charged vector field
in the bulk as a vector order parameter. The authors of Ref.
\cite{GubserPufu} proposed a holographic p-wave model by adding an
$SU(2)$ Yang-Mills field into the bulk, where a gauge boson
generated by one $SU(2)$ generator is dual to the vector order
parameter. Recently, Ref. \cite{CaiPWave-1} presented a new
holographic p-wave superconductor model by introducing a charged
vector field into an Einstein-Maxwell theory with a negative
cosmological constant. In the probe limit, the black hole solution
with non-trivial vector field can describe a superconducting phase
and the ratio of the gap frequency to the critical temperature is
given by $\omega_g/T_c\approx 8$, which is consistent with the
s-wave model \cite{HorowitzPRD78}. When taking the backreaction into
account, a rich phase structure: zeroth order, first order and
second order phase transitions has been observed in this p-wave
model \cite{CaiPWave-2,CaiPWave-3,CaiPWave-4}. Considering a
five-dimensional AdS soliton background coupled to a Maxwell complex
vector field, in Ref. \cite{CaiPWave-5} the authors reconstructed
the holographic p-wave insulator/superconductor phase transition
model in the probe limit and showed that the
Einstein-Maxwell-complex vector field model is a generalization of
the $SU(2)$ model with a general mass and gyromagnetic ratio. Other
generalized investigations based on this new p-wave model can be
found, for example, in Refs.
\cite{WuLuPRD,CaiYangPWave,WuLuPWaveIJMPA,ZPJ2015,CSJHEP2015,Nie2015,Rogatko2015}.
Considering the increasing interest in study of the holographic
p-wave model, we will also extend the study to the holographic
p-wave superconductor with hyperscaling violation, which has not
been constructed as far as we know. We will observe that the
hyperscaling violation has completely different effect on the phase
transitions for the holographic s-wave and p-wave superconductors,
and the Maxwell complex vector model is still a generalization of
the $SU(2)$ model even in the hyperscaling violation geometry. For
simplicity and clarity, in this work we will concentrate on the
probe limit where the backreaction of matter fields on the spacetime
metric is neglected.

The plan of the work is the following. In Sec. II we will briefly
review the black hole background with hyperscaling violation. In
Sec. III we will explore the effect of the hyperscaling violation on
the holographic s-wave superconductors. In Sec. IV we will discuss
the p-wave cases. We will conclude in the last section with our main
results. We will derive, in appendix A, the quations of motion for
the $SU(2)$ p-wave superconductors with hyperscaling violation.

\section{Black hole solution with hyperscaling violation}

In order to study the effect of hyperscaling violation on the
holographic superconductors in the probe limit, we consider the
black hole solution with hyperscaling violation \cite{DHKTW,ACYJHEP}
\begin{eqnarray}\label{Soliton}
ds_{d+2}^2=r^{-2(d-\theta)/d}\left[-r^{-2(z-1)}f(r)dt^2+\frac{dr^2}{f(r)}+dx_{i}^{2}\right],
\end{eqnarray}
with
\begin{eqnarray}\label{Metricf}
f(r)=1-\left(\frac{r}{r_{+}}\right)^{d+z-\theta},
\end{eqnarray}
where $\theta$ is the hyperscaling violation exponent, $z$ is the
dynamical exponent and $r_{+}$ is the radius of the event horizon.
Obviously, at the asymptotic boundary ($r\rightarrow0$), we have
\begin{eqnarray}\label{AsymptoticSoliton}
ds_{d+2}^2=r^{-2(d-\theta)/d}\left[-r^{-2(z-1)}dt^2+dr^2+dx_{i}^{2}\right],
\end{eqnarray}
which is the most general metric that is spatially homogeneous and
covariant under the scale transformations
\begin{eqnarray}\label{ScaleTransformation}
t\rightarrow\alpha^{z}t,~~(r,~x_{i})\rightarrow\alpha
(r,~x_{i}),~~ds\rightarrow\alpha^{\theta/d}ds,
\end{eqnarray}
with a real positive number $\alpha$. The novel feature of this
metric is that the proper distance $ds$ of the spacetime transforms
non-trivially under scale transformations with the exponent
$\theta$, which indicates a hyperscaling violation in the dual field
theory \cite{DHKTW,ACYJHEP}. Note that the Hawking temperature of
the black hole is determined by
\begin{eqnarray}\label{HawkingTemperature}
T=\frac{d+z-\theta}{4\pi r_{+}^{z}},
\end{eqnarray}
we can find that the thermal entropy, which is proportional to the
area of the black hole, becomes \cite{DHKTW}
\begin{eqnarray}\label{Entropy}
S\sim T^{(d-\theta)/z},
\end{eqnarray}
which establishes that $\theta$ is the hyperscaling violation
exponent. It should be noted that the metric (\ref{Soliton}) will
reduce to the pure Lifshitz case when $\theta=0$ and $z\neq1$, while
describe the pure AdS case when $\theta=0$ and $z=1$.

For convenience in the following discussion, we introduce the
coordinate transformation $u=r/r_{+}$ and rewrite the metric
(\ref{Soliton}) into
\begin{eqnarray}\label{SolitonU}
ds_{d+2}^2=(r_{+}u)^{-2(d-\theta)/d}\left[-(r_{+}u)^{-2(z-1)}f(u)dt^2+\frac{r_{+}^{2}}{f(u)}du^2+dx_{i}^{2}\right],
\end{eqnarray}
with $f(u)=1-u^{d+z-\theta}$. Considering the validity of the
solution and the requirement of the null energy condition, we get
the constraint \cite{DHKTW,ACYJHEP}
\begin{eqnarray}\label{Constraint}
d>\theta\geq0,~~z\geq1+\frac{\theta}{d}.
\end{eqnarray}
In this work, we will set $d=2,~z=2$ since we concentrate on the
effect of hyperscaling violation $\theta$ on the holographic
superconductors and compare with the results given in Ref.
\cite{FanJHEP}.

\section{S-wave superconductor models with hyperscaling violation}

In Ref. \cite{FanJHEP}, the author constructed the holographic
s-wave superconductors with hyperscaling violation and found that
the critical temperature increases as the hyperscaling violation
increases for the case of $z=2$ Lifshitz scaling. Now we will
reinvestigate the effect of the hyperscaling violation on the
holographic s-wave superconductors.

\subsection{Condensation and phase transition}

In the background of the four-dimensional hyperscaling violation
black hole, we consider a gauge field and a scalar field coupled via
the action
\begin{eqnarray}\label{System}
S=\int d^{4}x\sqrt{-g}\left(
-\frac{1}{4}F_{\mu\nu}F^{\mu\nu}-|\nabla\psi - iA\psi|^{2}
-m^2|\psi|^2 \right) \ .
\end{eqnarray}
Taking the ansatz of the matter fields as $\psi=\psi(u)$ and
$A=\phi(u) dt$, we can obtain the equations of motion from the
action (\ref{System}) for the scalar field $\psi$ and gauge field
$\phi$
\begin{eqnarray}
&&\psi^{\prime\prime}+\left(\frac{f^\prime}{f}-
\frac{3-\theta}{u}\right)\psi^\prime
+\left(\frac{r_{+}^{4}u^{2}\phi^2}{f^2}-\frac{m^2r_{+}^{\theta}}{u^{2-\theta}f}\right)\psi=0,
\label{BHPsiu}
\end{eqnarray}
\begin{eqnarray}
\phi^{\prime\prime}+\frac{1}{u}\phi^\prime-\frac{2r_{+}^{\theta}\psi^{2}}{u^{2-\theta}f}\phi=0,\label{BHPhiu}
\end{eqnarray}
where the prime denotes the derivative with respect to $u$.
Obviously, Eqs. (\ref{BHPsiu}) and  (\ref{BHPhiu}) reduce to ones in
the standard holographic s-wave superconductors with $z=2$ Lifshitz
scaling discussed in \cite{BuPRD} when $\theta\rightarrow0$. For
completeness, we will also present the results for the case
$\theta=0$ in the following.

In order to get the solutions in the superconducting phase, we have
to count on the appropriate boundary conditions for $\psi$ and
$\phi$. At the event horizon $u=1$ of the black hole, the regularity
gives the boundary conditions
\begin{eqnarray}
\psi^\prime(1)=-\frac{m^{2}r_{+}^{\theta}}{4-\theta}\psi(1)\,,\hspace{0.5cm}
\phi(1)=0\,. \label{horizon}
\end{eqnarray}
At the asymptotic boundary $u\rightarrow0$, the solutions behave
like
\begin{eqnarray}\label{PsiInfinity}
\psi=\left\{
\begin{array}{rl}
&\psi_{4-\Delta}r_{+}^{4-\Delta}u^{4-\Delta}+\psi_{\Delta}r_{+}^{\Delta}u^{\Delta}\,,~~~{\rm
with}\ \Delta=2+\sqrt{4+m^{2}}\ {\rm for}\ \theta=0,
\\ \\ &\psi_{0}+\psi_{\Delta}r_{+}^{\Delta}u^{\Delta}  \,,~~
\quad\quad\quad\quad\quad\quad {\rm with}\ \Delta=4-\theta\ {\rm
for}\ 0<\theta<2,
\end{array}\right.
\end{eqnarray}
\begin{eqnarray}
\phi=\rho+\mu\ln u\,. \label{PhiInfinity}
\end{eqnarray}
According to the gauge/gravity duality, $\Delta$ is the conformal
dimension of the scalar operator $\psi_{\Delta}=\langle O\rangle$
dual to the bulk scalar field, $\mu$ and $\rho$ are interpreted as
the chemical potential and charge density in the dual field theory
respectively. Just as in Refs. \cite{BuPRD,FanJHEP}, we will impose
boundary condition $\psi_{4-\Delta}=0$ and $\psi_{0}=0$ since we
focus on the condensate for the operator $\langle O\rangle$.

\subsubsection{Numerical investigation}

Using the shooting method \cite{HartnollPRLJHEP}, we can solve
numerically the equations of motion (\ref{BHPsiu}) and
(\ref{BHPhiu}) by doing integration from the horizon out to the
boundary even if the hyperscaling violation exponent $\theta$ is
fractional. Interestingly, from Eqs. (\ref{BHPsiu}) and
(\ref{BHPhiu}) we can get the useful scaling symmetry and induced
transformation of the relevant quantities
\begin{eqnarray}
&&\psi\rightarrow\alpha^{-\theta/2}\psi,\hspace{0.5cm}\phi\rightarrow\alpha^{-2}\phi,\nonumber\\
&&\psi_{\Delta}\rightarrow\alpha^{-(\Delta+\frac{\theta}{2})}\psi_{\Delta},\hspace{0.5cm}\mu\rightarrow\alpha^{-2}\mu,\hspace{0.5cm}
\label{SLsymmetry}
\end{eqnarray}
which can be used to build the invariant and dimensionless
quantities in the following calculation.

\begin{figure}[ht]
\includegraphics[scale=0.65]{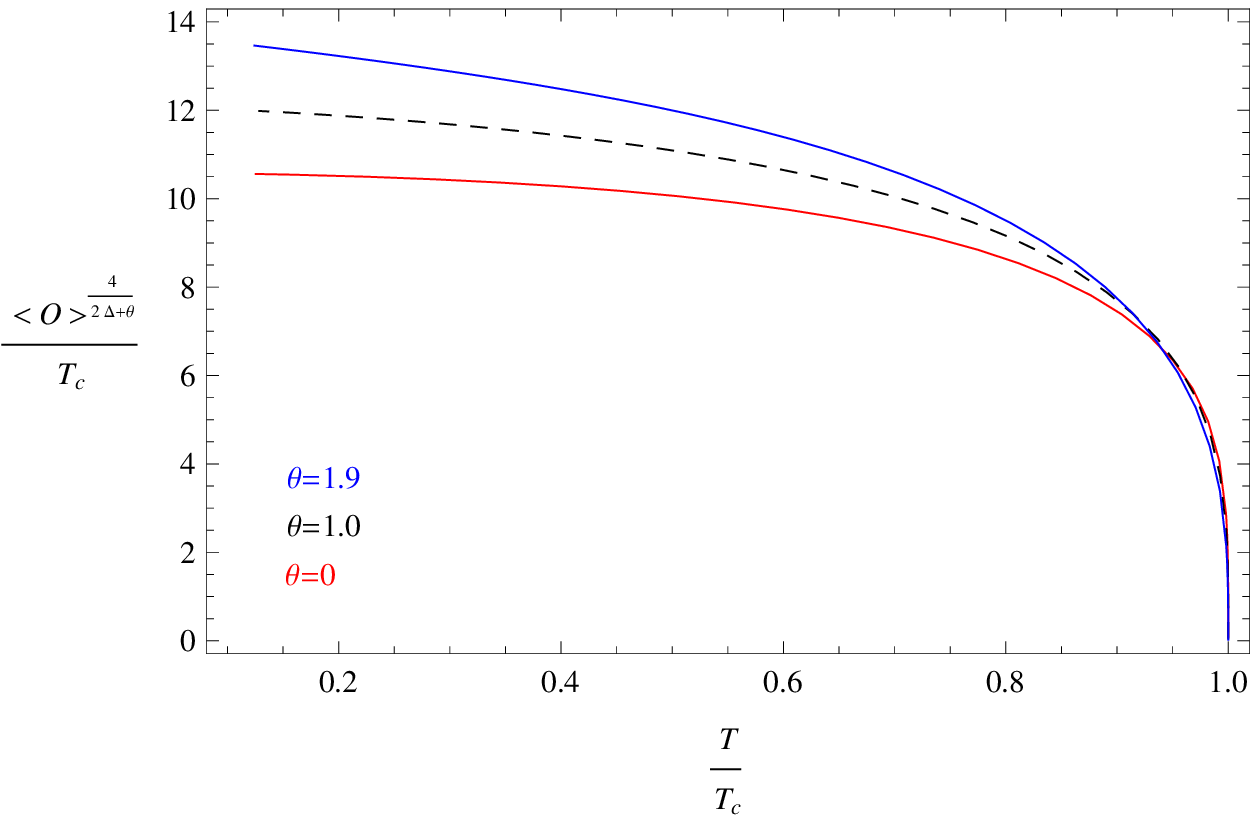}\hspace{0.2cm}%
\includegraphics[scale=0.65]{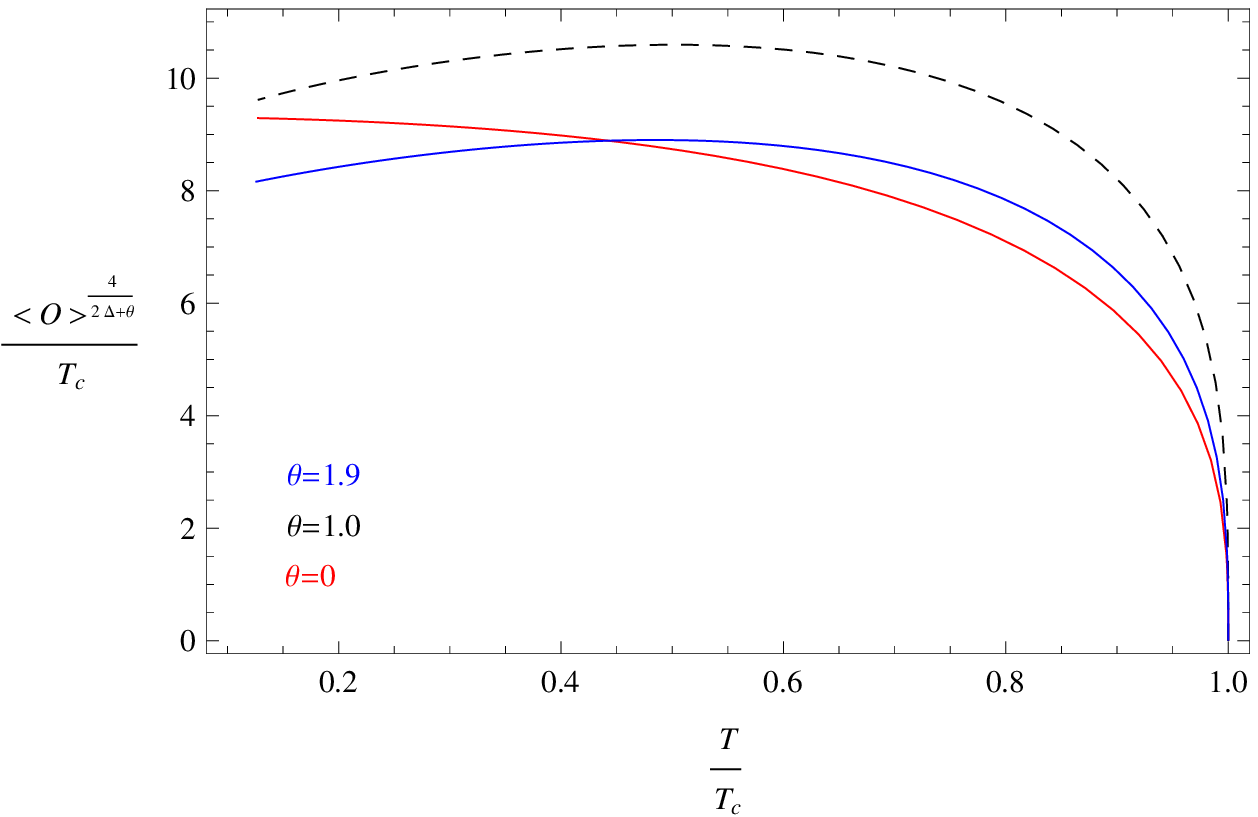}\\ \vspace{0.0cm}
\caption{\label{Condensate} (Color online) The condensate of the
scalar operator $\langle O\rangle$ as a function of temperature for
the fixed masses of the scalar field $m^2r_{+}^{\theta}=0$ (left)
and $m^2r_{+}^{\theta}=-3$ (right). In each panel, the three lines
correspond to different values of the hyperscaling violation, i.e.,
$\theta=0$ (red), $1.0$ (black and dashed) and $1.9$ (blue)
respectively.}
\end{figure}

Changing the hyperscaling violation $\theta$, we present in Fig.
\ref{Condensate} the condensate of the scalar operator $\langle
O\rangle$ as a function of temperature for the fixed masses of the
scalar field $m^2r_{+}^{\theta}=0$ (left) and $m^2r_{+}^{\theta}=-3$
(right). It is found that, for all cases considered here, the scalar
operator $\langle O\rangle$ is single-valued near the critical
temperature and the condensate drops to zero continuously at the
critical temperature. By fitting these curves, we see that for small
condensate there is a square root behavior
\begin{eqnarray}
\langle O\rangle\sim(1-T/T_{c})^{1/2}, \label{SquareRBehavior}
\end{eqnarray}
which is typical of second order phase transitions with the mean
field critical exponent $1/2$ for all values of $\theta$. The
behaviors of the condensate for the scalar operator $\langle
O\rangle$ show that the holographic s-wave superconductors still
exist even in the background of the hyperscaling violation black
hole.

\begin{table}[ht]
\begin{center}
\caption{\label{NMTcTable} The critical temperature $T_{c}$ obtained
by the numerical shooting method with the chosen various values of
the hyperscaling violation $\theta$ and scalar mass
$m^2r_{+}^{\theta}$ for the scalar operator $\langle O\rangle$ in
the holographic s-wave superconductor models with hyperscaling
violation. It is interesting to note that the mass of the scalar
field will not modify the value of the hyperscaling violation
$\theta_{*}\approx0.6$ which gives the minimum critical
temperature.}
\begin{tabular}{c c c c c c c c}
         \hline \hline
$\theta$ & 0 & 0.5 & 0.6 & 0.7 & 1.0 & 1.5 & 1.9
        \\
        \hline
~$m^2r_{+}^{\theta}=0$~~&~~$0.0229931\mu$~~&~~$0.0229413\mu$~~&~~$0.0229385\mu$~~~&~~$0.0229387\mu$
~~&~~$0.0229604\mu$~~&~~$0.0230883\mu$~~&~~$0.0233101\mu$~
          \\
~$m^2r_{+}^{\theta}=-1$~~&~~$0.0253991\mu$~~&~~$0.0251925\mu$~~&~~$0.0251835\mu$~~~&~~$0.0251853\mu$
~~&~~$0.0252606\mu$~~&~~$0.0256748\mu$~~&~~$0.0264006\mu$~
          \\
~$m^2r_{+}^{\theta}=-2$~~&~~$0.0289432\mu$~~&~~$0.0283204\mu$~~&~~$0.0282962\mu$~~~&~~$0.0283004\mu$
~~&~~$0.0284872\mu$~~&~~$0.0295178\mu$~~&~~$0.0314512\mu$~
          \\
~$m^2r_{+}^{\theta}=-3$~~&~~$0.0351935\mu$~~&~~$0.0330971\mu$~~&~~$0.0330311\mu$~~~&~~$0.0330385\mu$
~~&~~$0.0334913\mu$~~&~~$0.0361219\mu$~~&~~$0.0420384\mu$~
          \\
        \hline \hline
\end{tabular}
\end{center}
\end{table}

In order to obtain the effect of the hyperscaling violation on the
critical temperature $T_{c}$, we give the critical temperature
$T_{c}$ for the scalar operator $\langle O\rangle$ when we fix the
masses of the scalar field  $m^2r_{+}^{\theta}=0$, $-1$, $-2$ and
$-3$ for different hyperscaling violation exponent $\theta$ in Table
\ref{NMTcTable}. For the same mass of the scalar field, it is clear
that with the increase of the hyperscaling violation $\theta$, the
critical temperature $T_{c}$ decreases first and then increases,
which is different from the result obtained in Ref. \cite{FanJHEP}
where the critical temperature was shown to increase almost linearly
when $\theta$ increases. Obviously, from Table \ref{NMTcTable}, we
find that there exits a minimum value of the critical temperature at
$\theta_{*}\approx0.6$ for the fixed mass of the scalar field, which
implies that the higher hyperscaling violation makes it harder for
the condensation to form in the range $0\leq\theta<\theta_{*}$ but
easier in the range $\theta_{*}<\theta<2$. Interestingly, we observe
that the mass of the scalar field will not alter the value of
$\theta_{*}$. It should be noted that how the hyperscaling violation
works in the holographic s-wave superconductors is still an open
question.

\subsubsection{Analytical understanding}

Using the S-L method \cite{Siopsis}, Fan analytically explored the
effect of the hyperscaling violation on the s-wave superconducting
transition temperature and found that the critical temperature
increases with the increase of hyperscaling violation
\cite{FanJHEP}. Obviously, the analytical result given in Ref.
\cite{FanJHEP} is not in agreement with our numerical calculation
and some important information is missing. We will improve the S-L
method to get the the analytical result which is consistent with the
numerical calculation.

At the critical temperature $T_{c}$, the scalar field $\psi=0$.
Thus, near the critical point the equation of motion (\ref{BHPhiu})
for the gauge field $\phi$ reduces to
\begin{eqnarray}
\phi^{\prime\prime}+\frac{1}{u}\phi^\prime=0.\label{NESWPhiCritical}
\end{eqnarray}
Considering the boundary condition (\ref{horizon}) for $\phi$, we
can get the solution to Eq. (\ref{NESWPhiCritical})
\begin{eqnarray}
\phi(u)=\lambda r_{+c}^{-2}\ln u, \label{PhiSolution}
\end{eqnarray}
where we have set $\lambda=\mu r_{+c}^{2}$ with the radius of the
horizon at the critical point $r_{+c}$.

Defining a trial function $F(u)$ which matches the boundary behavior
(\ref{PsiInfinity}) for $\psi$
\begin{eqnarray}
\psi(u)\sim\langle O\rangle r_{+}^{\Delta}u^{\Delta}F(u),
\label{BintroduceF}
\end{eqnarray}
with the boundary conditions $F(0)=1$ and $F'(0)=0$, from Eq.
(\ref{BHPsiu}) we can obtain the equation of motion for $F(u)$
\begin{eqnarray}\label{BFEoM}
(QF^{\prime})^{\prime}+Q\left(U+\lambda^2V\right)F=0,
\end{eqnarray}
with
\begin{eqnarray}
Q(u)=u^{2\Delta+\theta-3}f,~~U(u)=\frac{\Delta
f'}{uf}+\frac{\Delta(\Delta+\theta-4)}{u^{2}}-\frac{m^2r_{+}^{\theta}}{u^{2-\theta}f},~~V(u)=\left(\frac{u\ln
u}{f}\right)^{2}.
\end{eqnarray}
According to the S-L eigenvalue problem \cite{Gelfand-Fomin}, we can
deduce the eigenvalue $\lambda$ minimizes the expression
\begin{eqnarray}\label{lambdaeigenvalue}
\lambda^{2}=\frac{\int^{1}_{0}Q\left(F'^{2}-UF^{2}\right)du}{\int^{1}_{0}QVF^2du}.
\end{eqnarray}
Using Eq. (\ref{lambdaeigenvalue}) to compute the minimum eigenvalue
of $\lambda^{2}$, we can get the critical temperature $T_{c}$ for
different hyperscaling violation $\theta$ and mass of the scalar
field $m^2r_{+}^{\theta}$ from the following relation
\begin{eqnarray}\label{CTTc}
T_{c}=\frac{4-\theta}{4\pi\lambda_{min}}\mu.
\end{eqnarray}

Before going further, we would like to give a comment. Considering
the boundary conditions of $F(u)$, i.e., $F(0)=1$ and $F'(0)=0$,
people usually assume the trial function to be
\begin{eqnarray}\label{FunctionFua}
F(u)=F_{a}(u)=1-au^{2},
\end{eqnarray}
with a constant $a$, just as done in Ref. \cite{FanJHEP}. It should
be noted that this assumption works well in most cases
\cite{Siopsis,Li-Cai-Zhang,ZengSL,JingJHEP2011,SDSL2012,PanJWCh,BGRLPRD2013,
GangopadhyayPLB,KuangPRD,LinEPJC,Nakonieczny2014,Lai2015,Nakonieczny2015}.
Unfortunately, using this trial function in our case, we find that
the analytical results are not in agreement with the numerical
calculation and some important information is missing. To solve this
problem, we choose the trial function by including the higher order
of $u$ such as the third order trial function
\begin{eqnarray}\label{FunctionFuab}
F(u)=F_{ab}(u)=1-au^{2}+bu^{3},
\end{eqnarray}
with two constants $a$ and $b$. We prefer (\ref{FunctionFuab}) over
(\ref{FunctionFua}) because it gives a better estimate of the
minimum of (\ref{lambdaeigenvalue}), which means that the analytical
results are much more closer to the numerical findings, and more
importantly, can rightly reveal the influence of the hyperscaling
violation on the critical temperature $T_{c}$.

\begin{figure}[ht]
\includegraphics[scale=0.65]{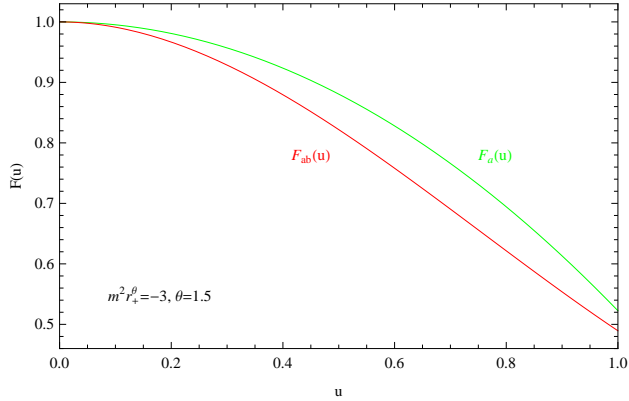}\\ \vspace{0.0cm}
\caption{\label{TrialFunction} (Color online) The trial function
$F_{a}(u)=1-au^{2}$ (green) and $F_{ab}(u)=1-au^{2}+bu^{3}$ (red) as
a function of $u$ when $\lambda^{2}$ attains its minimum for the
fixed mass of the scalar field $m^2r_{+}^{\theta}=-3$ and
hyperscaling violation $\theta=1.5$.}
\end{figure}

As an example, we will calculate the case for the fixed mass of the
scalar field $m^2r_{+}^{\theta}=-3$ with the chosen value of the
hyperscaling violation $\theta=1.5$. If we choose the trial function
(\ref{FunctionFua}), we have
\begin{eqnarray}
\lambda^{2}=\lambda_{a}^{2}=\frac{0.5-0.785714a+0.490385a^2}{0.0137786-0.0147955a+0.00459295a^2},
\end{eqnarray}
whose minimum is $\lambda^{2}=30.4899$ at $a=0.477530$. According to
Eq. (\ref{lambdaeigenvalue}), we can easily obtain the critical
temperature $T_{c}=0.036029\mu$. Changing the trial function into
the form (\ref{FunctionFuab}), we arrive at
\begin{eqnarray}
\lambda^{2}=\lambda_{ab}^{2}=\frac{0.5-0.785714a+0.490385a^2-0.983333ab+0.705357b+0.508824b^2}
{0.0137786-0.0147955a+0.00459295a^2-0.00750497ab+0.0114973b+0.00312254b^2},
\end{eqnarray}
whose minimum is $\lambda^{2}=30.3335$ at $a=0.913984$ and
$b=0.403370$. Hence the critical temperature reads
$T_{c}=0.0361217\mu$. Comparing with the analytical result from the
trial function $F_{a}(u)$, we find that this value is much more
closer to the numerical result $T_{c}=0.0361219\mu$ given in Table
\ref{NMTcTable} and the difference between the analytical and
numerical values is $0.0004\%$! In Fig. \ref{TrialFunction}, we plot
the trial function $F_{a}(u)=1-au^{2}$ (green) and
$F_{ab}(u)=1-au^{2}+bu^{3}$ (red) as a function of $u$ when
$\lambda^{2}$ attains its minimum for the fixed mass of the scalar
field $m^2r_{+}^{\theta}=-3$ and hyperscaling violation
$\theta=1.5$, which clearly shows the difference between these two
trial functions.

\begin{figure}[ht]
\includegraphics[scale=0.65]{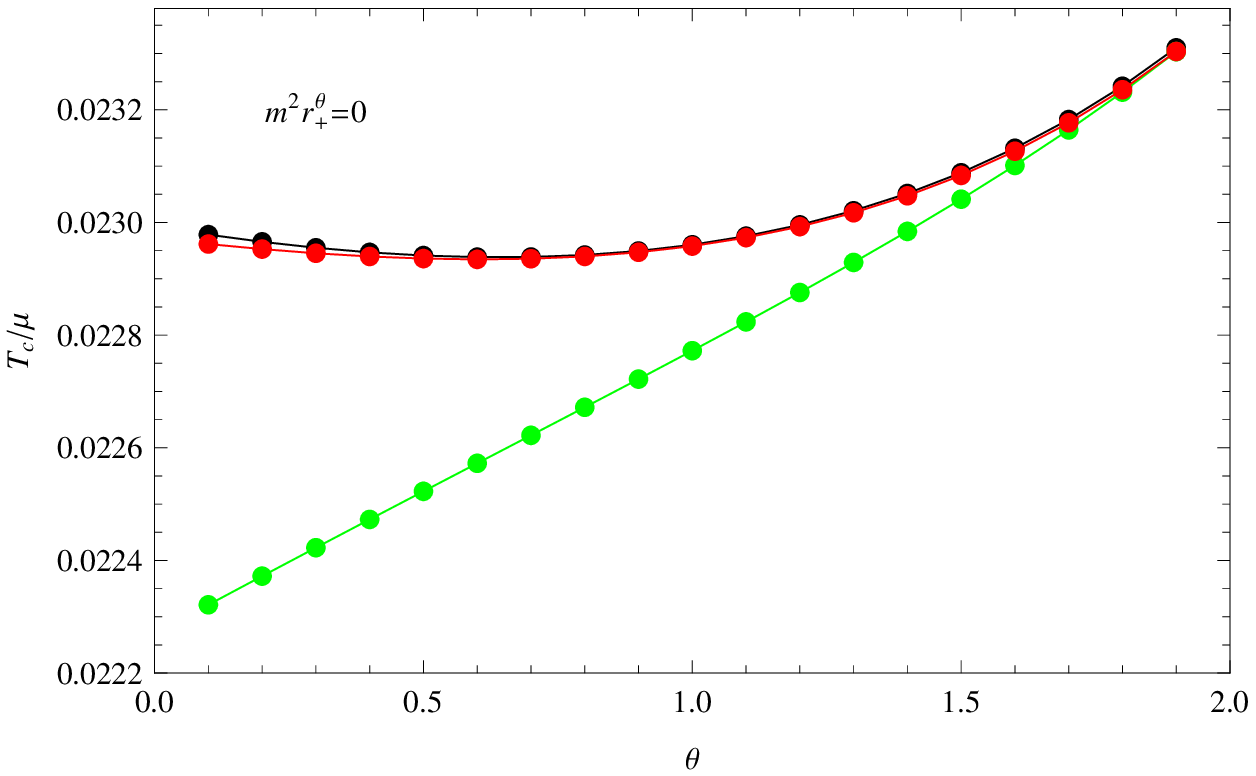}\hspace{0.2cm}%
\includegraphics[scale=0.65]{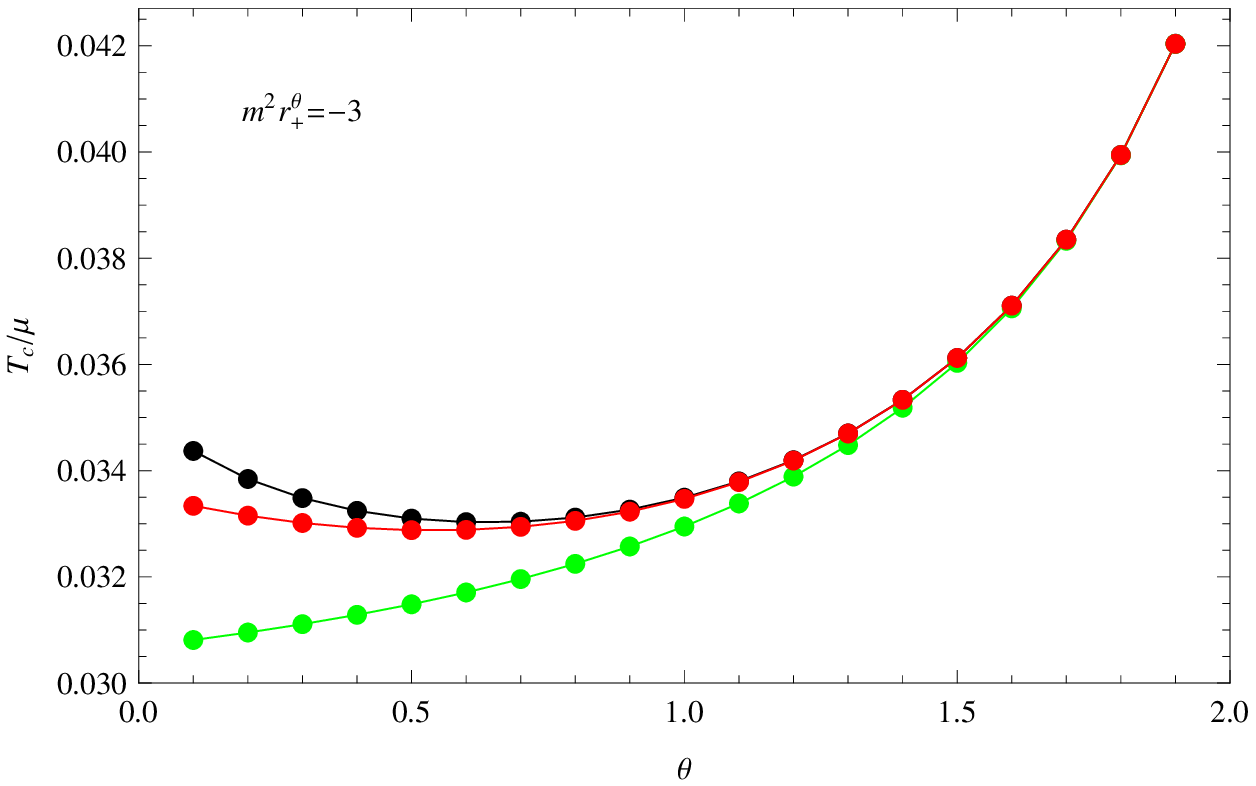}\\ \vspace{0.0cm}
\caption{\label{CriticalTemperature} (Color online) The critical
temperature $T_{c}$ as a function of the hyperscaling violation
$\theta$ with fixed masses of the scalar field
$m^{2}r_{+}^{\theta}=0$ (left) and $m^{2}r_{+}^{\theta}=-3$ (right)
in the holographic s-wave superconductors with hyperscaling
violation. The three lines from top to bottom correspond to the
results obtained by the numerical calculation (black) and from the
analytical S-L method by using the trial function
$F(u)=1-au^{2}+bu^{3}$ (red) and $F(u)=1-au^{2}$ (green),
respectively.}
\end{figure}

Extending the analytical investigation to the holographic s-wave
superconductors with hyperscaling violation in the full parameter
space (\ref{Constraint}), we can get the critical temperature
$T_{c}$ with various masses of the scalar field $m^2r_{+}^{\theta}$
and hyperscaling violation $\theta$ for the scalar operator $\langle
O\rangle$. To see the dependence of the analytical results on the
hyperscaling violation more directly and compare with Figure 3 of
Ref. \cite{FanJHEP}, we exhibit the critical temperature $T_{c}$ as
a function of the hyperscaling violation $\theta$ for the fixed
masses of the scalar field $m^2r_{+}^{\theta}=0$ (left) and
$m^2r_{+}^{\theta}=-3$ (right) in Fig. \ref{CriticalTemperature}. We
also present the numerical results obtained by using the shooting
method in order to compare with the analytical results. Obviously,
compared with the trial function $F_{a}(u)$, the third order trial
function $F_{ab}(u)$ can indeed be used to further improve the
analytical results and reduce the disparity between the analytical
and numerical results. And what's more, in contrast to the trial
function $F_{a}(u)$ which only tells us that the critical
temperature $T_{c}$ increases with the increase of hyperscaling
violation $\theta$, the third order trial function $F_{ab}(u)$ can
be used to give the analytical results which are completely
consistent with the numerical findings. Thus, we conclude that we
can still count on the S-L method with the higher order trial
function $F(u)$ to analytically study the effect of the hyperscaling
violation on the holographic s-wave superconductors.

Now we are in a position to study the critical phenomena of the
holographic s-wave superconductors with hyperscaling violation.
Since the condensation for the scalar operator $\langle O\rangle$ is
so small when $T \rightarrow T_c$, we can expand $\phi(u)$ in
$\langle O\rangle$ as
\begin{eqnarray}\label{PhiExpandNearTc}
\phi(u)=\lambda r_{+}^{-2}\ln
u+r_{+}^{-2}\mathcal{A}\chi(u)+\cdot\cdot\cdot,
\end{eqnarray}
where we have introduced $\mathcal{A}=r_{+}^{2\Delta+\theta}\langle
O\rangle^{2}$ and the boundary conditions $\chi(1)=0$ and
$\chi'(1)=0$ \cite{Siopsis}. Thus, substituting the functions
(\ref{BintroduceF}) and (\ref{PhiExpandNearTc}) into (\ref{BHPhiu}),
we can get the equation of motion for $\chi(u)$
\begin{eqnarray}\label{BHChiuEoM}
(u\chi^\prime)^\prime=\frac{2\lambda u^{2\Delta+\theta-1}F^{2}\ln
u}{f}.
\end{eqnarray}
Making integration of both sides of Eq. (\ref{BHChiuEoM}), we have
\begin{eqnarray}\label{Chi0}
(u\chi^\prime)|_{u\rightarrow
0}=\lambda\mathcal{C}=-\lambda\int^{1}_{0}\frac{2u^{2\Delta+\theta-1}F^{2}\ln
u}{f}du.
\end{eqnarray}

From Eqs. (\ref{PhiSolution}) and (\ref{PhiExpandNearTc}), near
$u\rightarrow0$ we can arrive at
\begin{eqnarray}\label{BHMuExp}
\mu=\lambda
r_{+}^{-2}+r_{+}^{-2}\mathcal{A}(u\chi^\prime)|_{u\rightarrow 0},
\end{eqnarray}
which leads to
\begin{eqnarray}\label{OExp}
\langle O\rangle=\frac{1}{\sqrt{\mathcal{C}}}\left(\frac{4\pi
T_{c}}{4-\theta}\right)^{\frac{2\Delta+\theta}{4}}\left(1-\frac{T}{T_c}\right)^{\frac{1}{2}}.
\end{eqnarray}
Obviously, the expression (\ref{OExp}) is valid for all cases
considered here. For example, for the case of $\theta=1.5$ with
$m^2r_{+}^{\theta}=-3$, we have $\langle
O\rangle=0.269906(1-T/T_c)^{1/2}$ when $a=0.913984$ and
$b=0.403370$, which agrees well with the numerical calculation by
using the shooting method. Especially, for the case of $\theta=0$
with $m^2=-3$, we obtain $\langle O\rangle=0.249154(1-T/T_c)^{1/2}$
when $a=1.64069$ and $b=0.911971$, which is in good agreement with
the numerical result given in \cite{BuPRD}. Since the hyperscaling
violation and mass of the scalar field will not alter Eq.
(\ref{OExp}) except for the prefactor, we can reproduce the
expression (\ref{SquareRBehavior}) near the critical point by using
the analytical S-L method. It shows that the holographic s-wave
superconducting phase transition with hyperscaling violation is of
the second order and the critical exponent of the system always
takes the mean-field value $1/2$. The hyperscaling violation will
not influence the result.

\subsection{Conductivity}

In \cite{FanJHEP}, it was found that a gap opens in the real part of
the conductivity in the holographic s-wave superconductor models
with hyperscaling violation, which indicates the onset of
superconductivity. Considering that the author only concentrated on
the case of the hyperscaling violation $\theta=1.0$, we will vary
$\theta$ to discuss the effect of the hyperscaling violation on the
conductivity.

Assuming that the perturbed Maxwell field has a form $\delta
A_{x}=A_{x}(u)e^{-i\omega t}dx$, we obtain the equation of motion
for $A_{x}$ which can be used to calculate the conductivity
\begin{eqnarray}
A_{x}^{\prime\prime}+\left(\frac{f^\prime}{f}-\frac{1}{u}\right)A_{x}^\prime
+\left(\frac{r_{+}^{4}\omega^2u^{2}}{f^2}-\frac{2r_{+}^{\theta}\psi^{2}}{u^{2-\theta}f}\right)A_{x}=0.
\label{ConductivityEquation}
\end{eqnarray}
For different hyperscaling violation exponents, the ingoing wave
boundary condition near the horizon is given by
\begin{eqnarray}
A_{x}(u)\sim (1-u)^{-\frac{i\omega}{4\pi T}},
\end{eqnarray}
and the general behavior in the asymptotic region ($u\rightarrow 0$)
can be written as
\begin{eqnarray}
A_{x}=A_{x}^{(0)}+A_{x}^{(1)}r_{+}^{2}u^{2}.
\end{eqnarray}
Thus, we can obtain the conductivity of the dual superconductor by
using the gauge/gravity duality \cite{HartnollPRLJHEP,FanJHEP}
\begin{eqnarray}\label{CMFConductivity}
\sigma=-\frac{2iA_{x}^{(1)}}{\omega A_{x}^{(0)}}\ .
\end{eqnarray}
For different values of the hyperscaling violation $\theta$, one can
obtain the conductivity by solving the Maxwell equation numerically.
We still focus on the case for the fixed scalar masses
$m^2r_{+}^{\theta}=0$ and $m^2r_{+}^{\theta}=-3$ in our discussion.

\begin{figure}[H]
\includegraphics[scale=0.425]{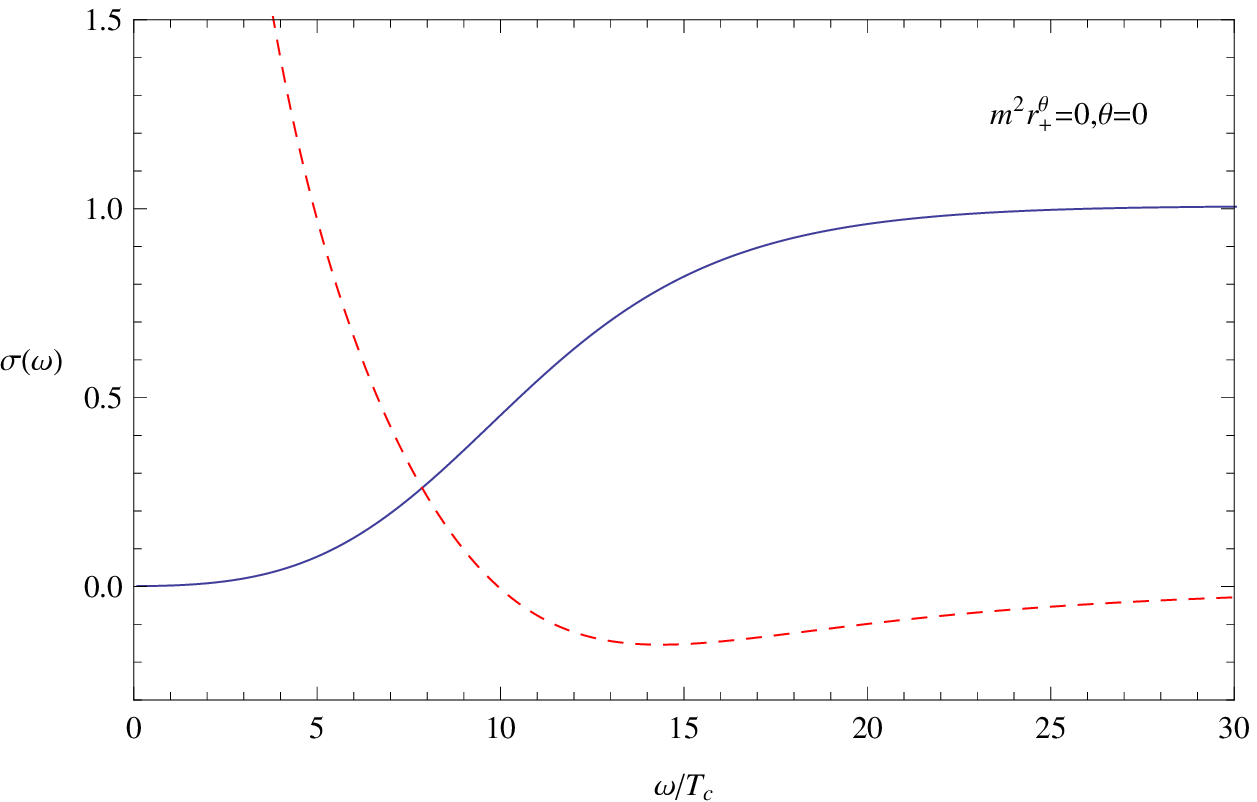}\hspace{0.2cm}%
\includegraphics[scale=0.425]{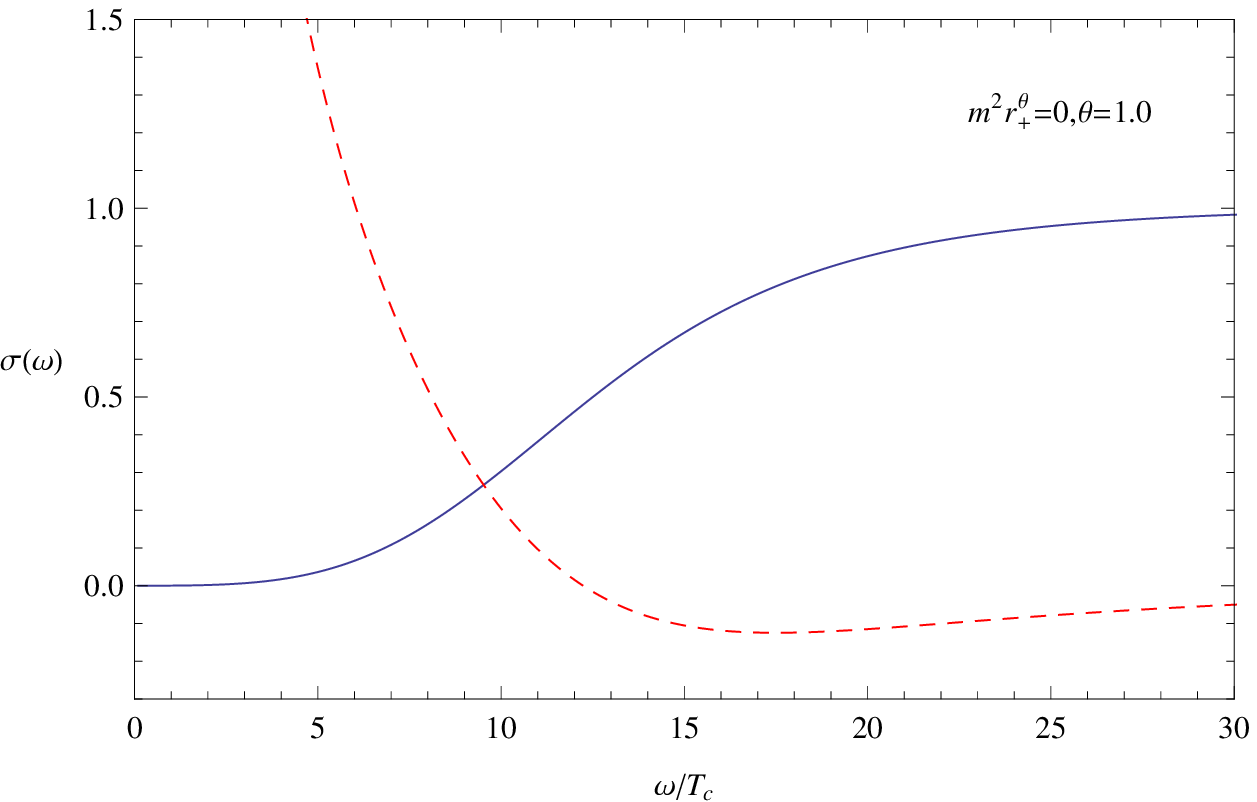}\hspace{0.2cm}%
\includegraphics[scale=0.425]{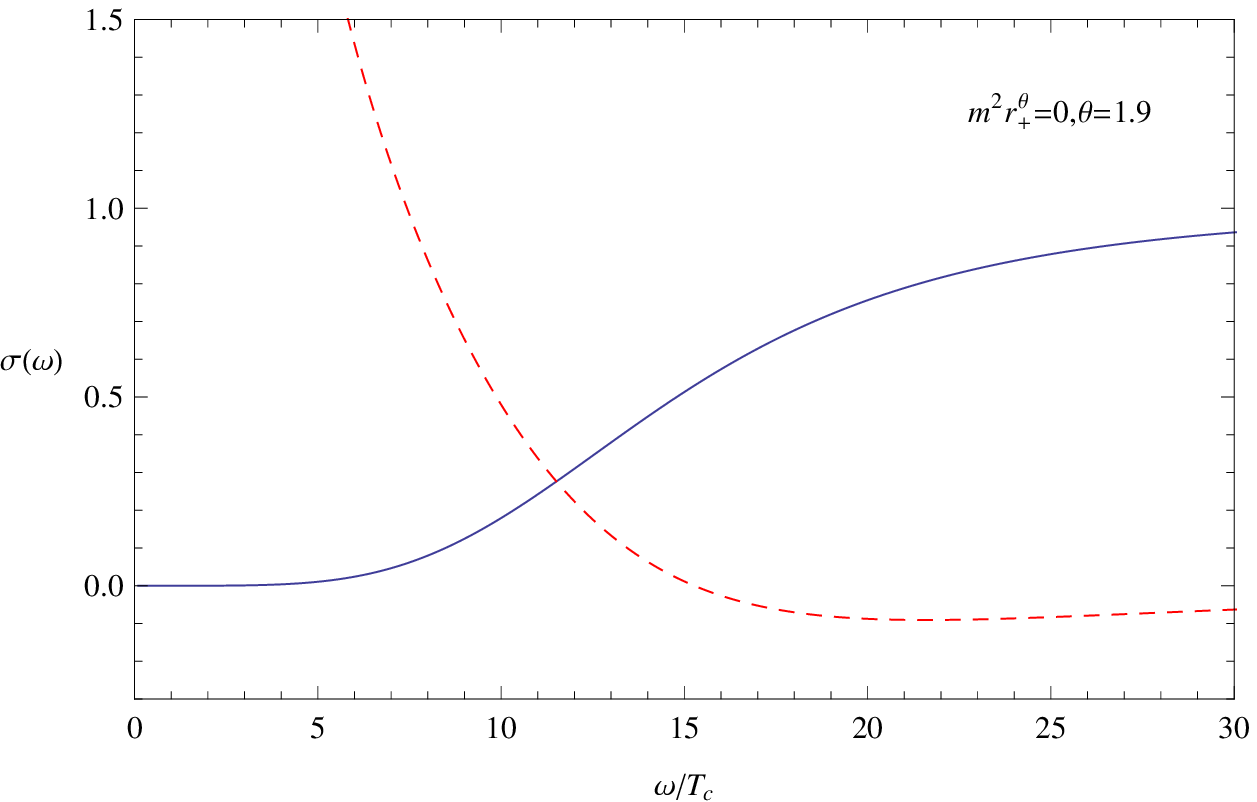}\\ \vspace{0.0cm}
\includegraphics[scale=0.425]{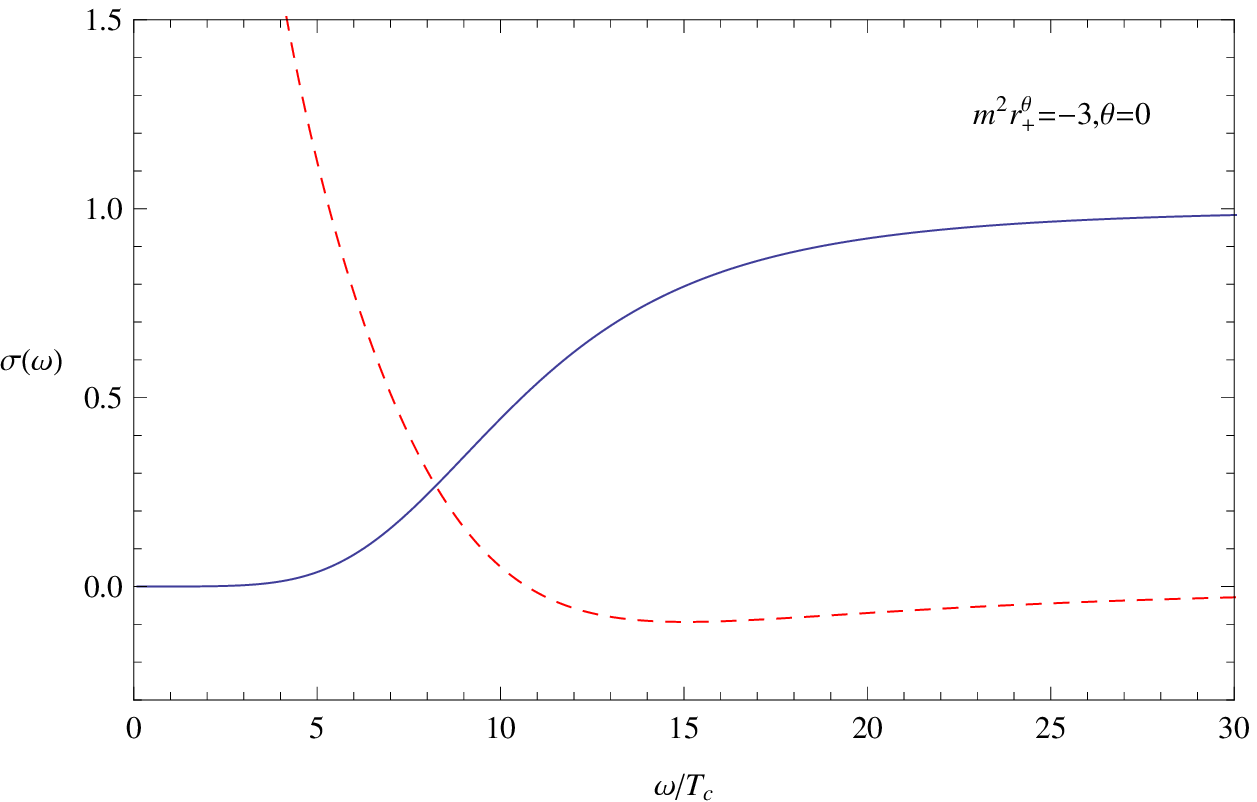}\hspace{0.2cm}%
\includegraphics[scale=0.425]{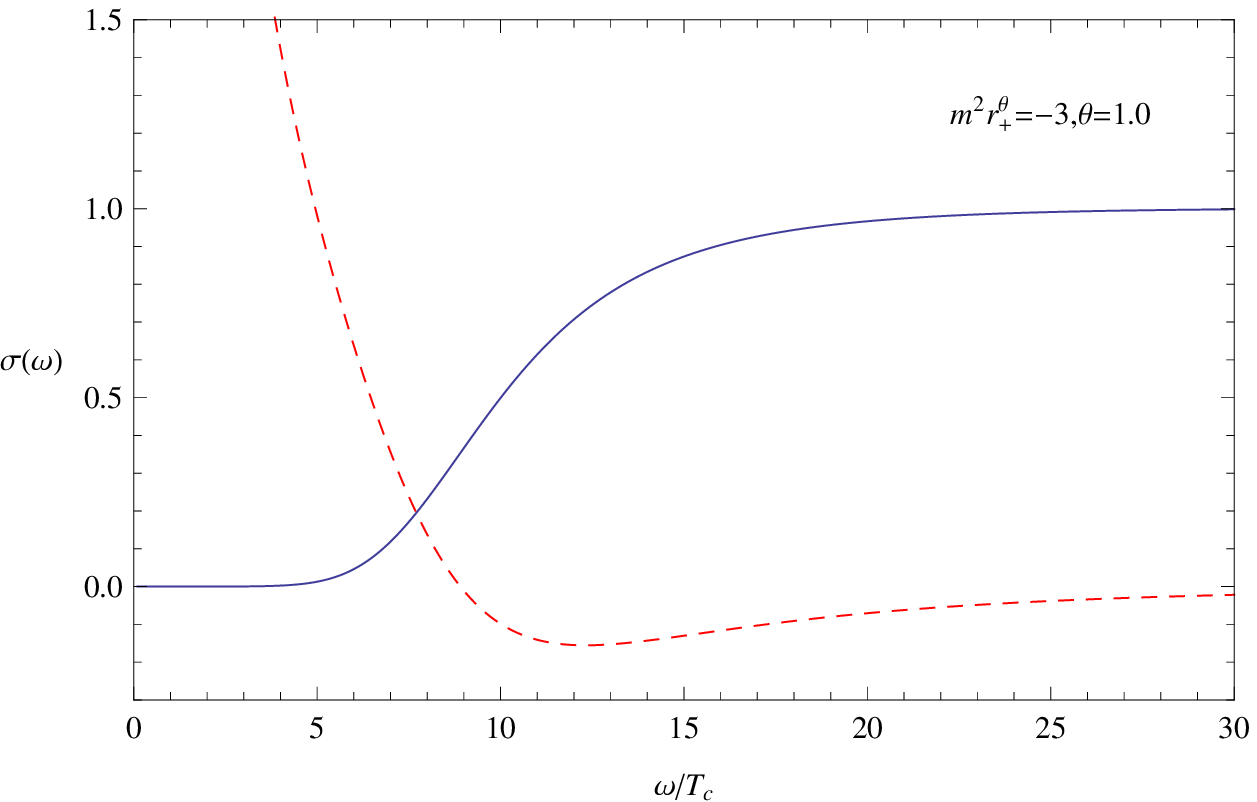}\hspace{0.2cm}%
\includegraphics[scale=0.425]{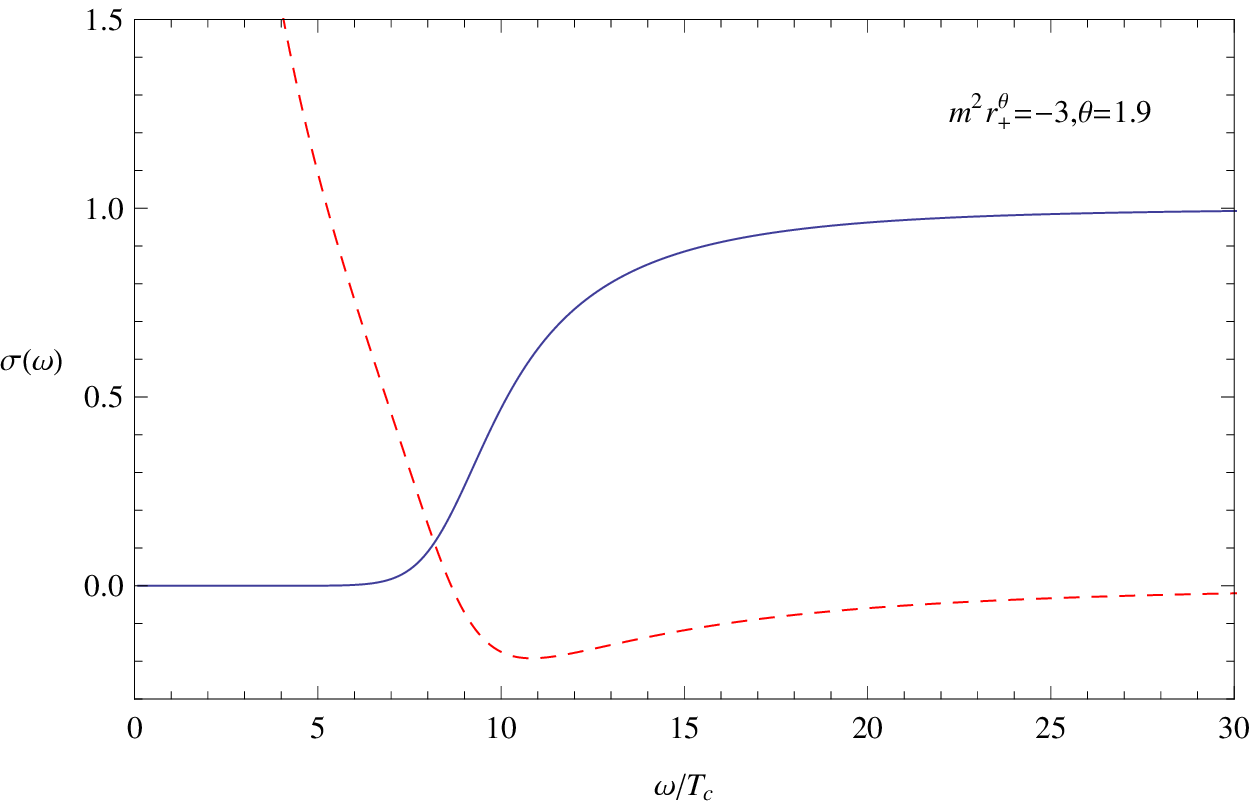}\\ \vspace{0.0cm}
\caption{\label{Conductivity} (color online) Conductivity of the
holographic s-wave superconductors with hyperscaling violation for
the fixed masses of the scalar field and different values of the
hyperscaling violation. In each panel, the blue (solid) line and red
(dashed) line represent the real part and imaginary part of the
conductivity $\sigma(\omega)$ respectively.}
\end{figure}

In Fig. \ref{Conductivity} we plot the frequency dependent
conductivity of the holographic s-wave superconductors with
hyperscaling violation by solving the Maxwell equation
(\ref{ConductivityEquation}) numerically for $\theta=0$, $1.0$ and
$1.9$ with $m^2r_{+}^{\theta}=0$ and $m^2r_{+}^{\theta}=-3$ at
temperatures $T/T_{c}\approx0.15$. In each panel, the blue (solid)
line and red (dashed) line represent the real part and imaginary
part of the conductivity $\sigma(\omega)$ respectively. For all
cases considered here, we find a gap in the conductivity which is
parameterized by the gap frequency $\omega_{g}$. Defining
$\omega_{g}$ by the minimum of $|\sigma|$ \cite{HorowitzPRD78}, we
observe that with the increase of the hyperscaling violation
$\theta$, the gap frequency $\omega_{g}$ becomes larger.  Also, for
increasing hyperscaling violation, we have larger deviation from the
value $\omega_g/T_c\approx 8$, especially for the case of
$m^2r_{+}^{\theta}=0$. This shows that the hyperscaling violation
really affects the conductivity of the holographic s-wave
superconductors and change the ratio in the gap frequency
$\omega_g/T_c\approx 8$ which was claimed to be universal in
\cite{HorowitzPRD78}.

\section{P-wave superconductor models with hyperscaling violation}

Since the hyperscaling violation has the interesting effect on the
holographic s-wave superconductors, which greatly improves the
previous findings in Ref. \cite{FanJHEP}, it seems worthwhile to
consider the influence of the hyperscaling violation on the
holographic p-wave superconductors which has not been constructed as
far as we know.

\subsection{Condensation and phase transition}

Working in the probe limit, we will construct the p-wave holographic
superconductor models with hyperscaling violation via the Maxwell
complex vector field model \cite{CaiPWave-1,CaiPWave-2}
\begin{eqnarray}\label{NewPWction}
S=\int
d^{4}x\sqrt{-g}\left(-\frac{1}{4}F_{\mu\nu}F^{\mu\nu}-\frac{1}{2}\rho_{\mu\nu}^{\dag}\rho^{\mu\nu}-m^2\rho_{\mu}^{\dag}\rho^{\mu}+i
q\gamma\rho_{\mu}\rho_{\nu}^{\dag}F^{\mu\nu}\right),
\end{eqnarray}
where the strength of $U(1)$ field $A_\mu$ is
$F_{\mu\nu}=\nabla_{\mu}A_{\nu}-\nabla_{\nu}A_{\mu}$ and the tensor
$\rho_{\mu\nu}$ is defined by
$\rho_{\mu\nu}=D_\mu\rho_\nu-D_\nu\rho_\mu$ with the covariant
derivative $D_\mu=\nabla_\mu-iqA_\mu$. $m$ and $q$ represent the
mass and charge of the vector field $\rho_\mu$, respectively. Since
we will consider the case without external magnetic field, the
parameter $\gamma$, which describes the interaction between the
vector field $\rho_\mu$ and gauge field $A_\mu$, will not play any
role.

Just as in Refs. \cite{CaiPWave-1,CaiPWave-2}, we will adopt the
following ansatz for the matter fields
\begin{eqnarray}\label{Ansatz}
\rho_{\nu}dx^{\nu}=\rho_{x}(u)dx\,,\hspace{0.5cm}A_{\nu}dx^{\nu}=\phi(u)dt,
\end{eqnarray}
where we can set $\rho_{x}$ to be real by using the $U(1)$ gauge
symmetry. Thus, from the solution (\ref{SolitonU}) for $d=2,~z=2$
and action (\ref{NewPWction}), we can get the equations of motion
for the vector hair $\rho_{x}$ and gauge field $\phi$
\begin{eqnarray}
\rho_{x}^{\prime\prime}+\left(\frac{f^\prime}{f}-\frac{1}{u}\right)\rho_{x}^\prime
+\left(\frac{r^{4}_{+}u^{2}\phi^2}{f^{2}}-\frac{m^2r_{+}^{\theta}}{u^{2-\theta}f}\right)\rho_{x}=0\,,
\label{NewPWRhox}
\end{eqnarray}
\begin{eqnarray}
\phi^{\prime\prime}+\frac{1}{u}\phi^\prime-\frac{2r_{+}^{2}\rho_x^2}{f}\phi=0,
\label{NewPWPhi}
\end{eqnarray}
where the prime denotes the derivative with respect to $u$. Without
loss of generality, we have scaled $q=1$ as in Ref.
\cite{CaiPWave-1}. Comparing the above two equations of motion with
Eqs. (\ref{pYMwave-Psi}) and (\ref{pYMwave-Phi}) for the holographic
p-wave superconductors with hyperscaling violation in the $SU(2)$
Yang-Mills system constructed in the appendix, we can find that the
two sets of equations of motion are the same if we set
$m^2r_{+}^{\theta}=0$ and rescale the field by
$\rho_{x}(u)=\psi(u)/\sqrt{2}$ in Eqs. (\ref{NewPWRhox}) and
(\ref{NewPWPhi}). Thus, in this sense, the complex vector field
model is still a generalization of the $SU(2)$ Yang-Mills model in
the holographic superconductors with hyperscaling violation, which
supports the argument given in \cite{CaiPWave-5}.

We will impose the appropriate boundary conditions for $\rho_{x}(u)$
and $\phi(u)$ to get the solutions in the superconducting phase.
Interestingly, we observe that $\phi$ has the same boundary
conditions just as in Eq. (\ref{horizon}) for the horizon $u=1$ and
Eq. (\ref{PhiInfinity}) for the boundary $u\rightarrow 0$. But for
the vector field $\rho_{x}$, we find at the horizon
\begin{eqnarray}
\rho_{x}^\prime(1)=-\frac{m^2r_{+}^{\theta}}{4-\theta}\rho_{x}(1)\,,
\label{RhoxHorizon}
\end{eqnarray}
and at the asymptotic boundary
\begin{eqnarray}\label{RhoxPhiInfinity}
\rho_{x}=\left\{
\begin{array}{rl}
&\rho_{x}^{(2-\Delta)}r_{+}^{2-\Delta}u^{2-\Delta}+\rho_{x}^{(\Delta)}r_{+}^{\Delta}u^{\Delta}\,,~~~~~{\rm
with}\ \Delta=1+\sqrt{1+m^{2}}\ {\rm for}\ \theta=0,
\\ \\ &\rho_{x}^{(0)}+\rho_{x}^{(\Delta)}r_{+}^{\Delta}u^{\Delta}\,,~~
\quad\quad\quad\quad\quad\quad\quad  {\rm with}\ \Delta=2\ {\rm
for}\ 0<\theta<2,
\end{array}\right.
\end{eqnarray}
where $\rho_{x}^{(2-\Delta)}$ (or $\rho_{x}^{(0)}$) and
$\rho_{x}^{(\Delta)}$ are interpreted as the source and vacuum
expectation value of the vector operator $J_{x}$ in the dual field
theory according to the gauge/gravity duality respectively. In this
work, we will impose boundary condition $\rho_{x}^{(2-\Delta)}=0$
(or $\rho_{x}^{(0)}=0$) since we require that the condensate appears
spontaneously.

We will count on the shooting method \cite{HartnollPRLJHEP} to solve
numerically the equations of motion (\ref{NewPWRhox}) and
(\ref{NewPWPhi}) in this section. In the following calculation, we
will use the scaling symmetry from the equations of motion and
induced transformation of the relevant quantities, i.e.,
\begin{eqnarray}
&&\rho_{x}\rightarrow\alpha^{-1}\rho_{x},\hspace{0.5cm}\phi\rightarrow\alpha^{-2}\phi,\nonumber\\
&&\rho_{x}^{(\Delta)}\rightarrow\alpha^{-(1+\Delta)}\rho_{x}^{(\Delta)},
\hspace{0.5cm}\mu\rightarrow\alpha^{-2}\mu,\hspace{0.5cm}
\label{PWSymmetry}
\end{eqnarray}
to build the invariant and dimensionless quantities.

\begin{figure}[ht]
\includegraphics[scale=0.65]{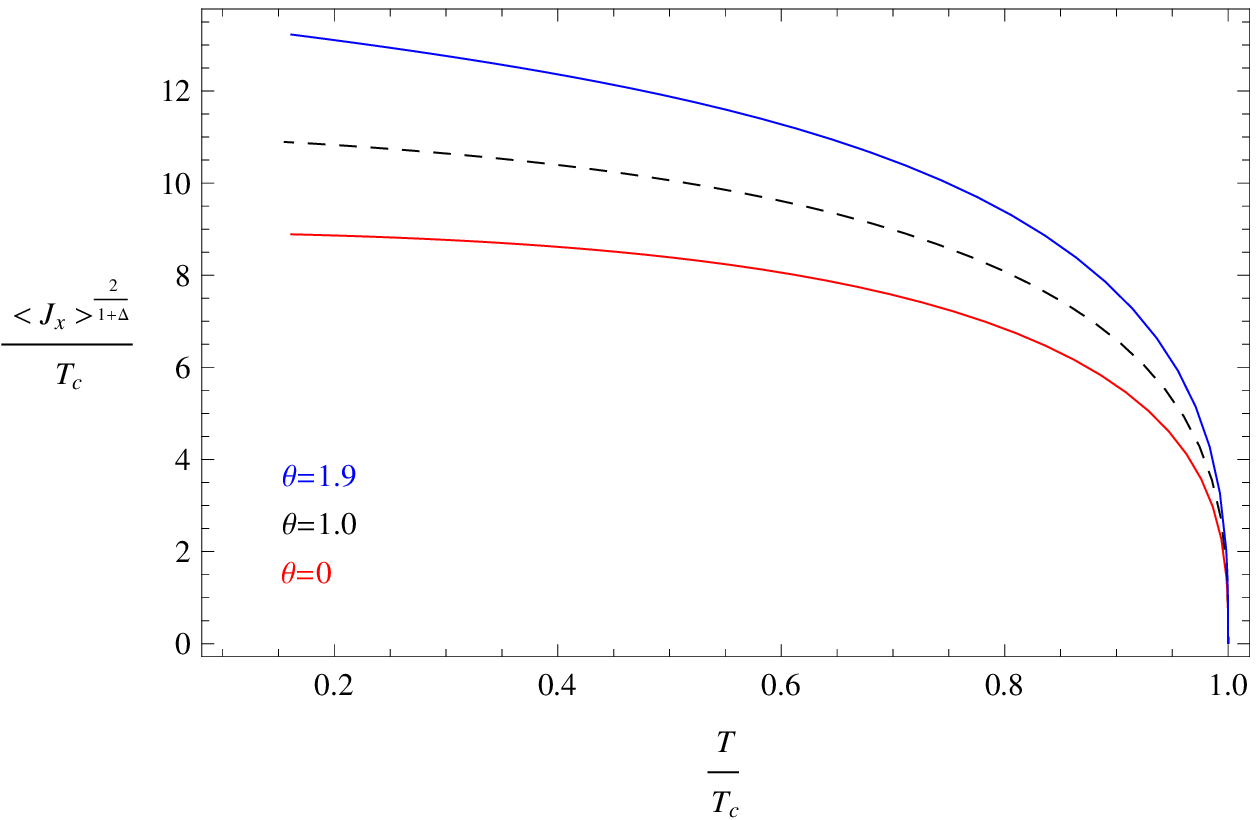}\vspace{0.0cm}
\includegraphics[scale=0.65]{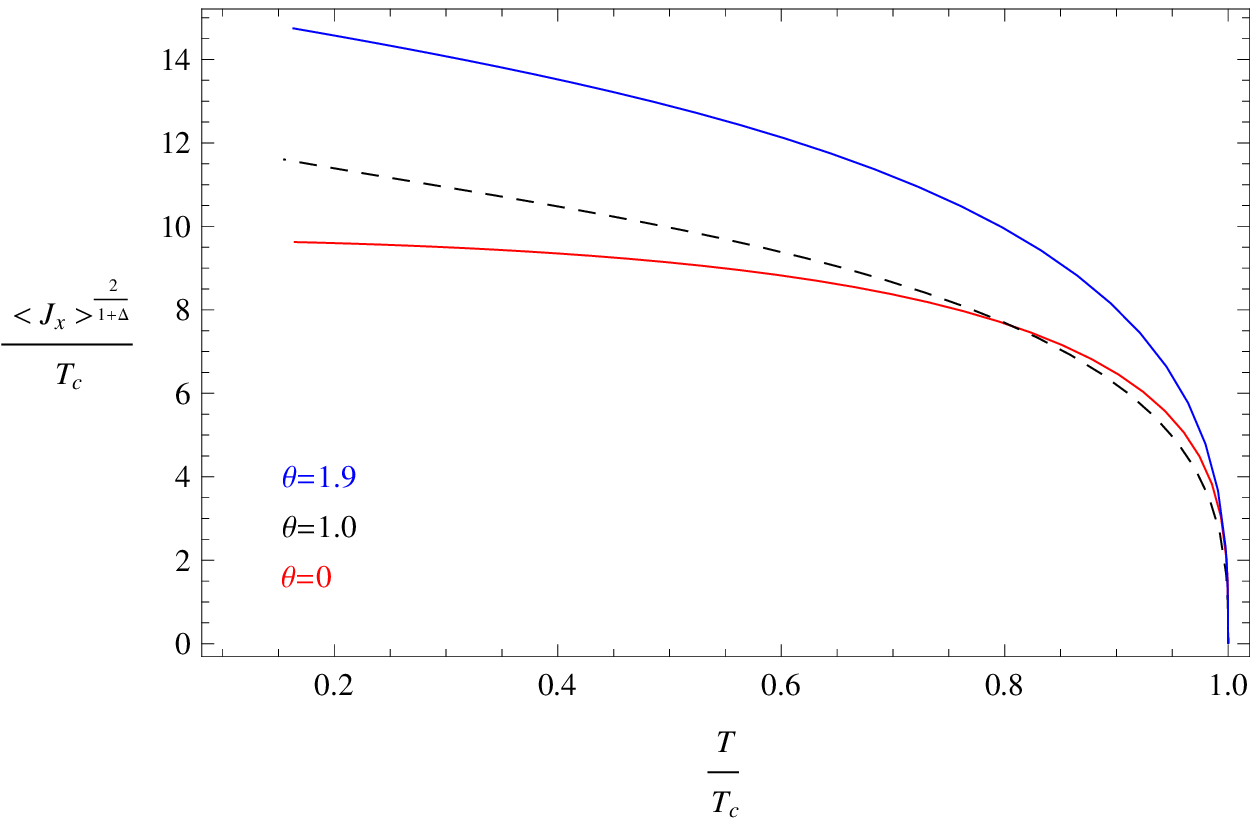}\\ \vspace{0.0cm}
\caption{\label{PWaveCond} (color online) The condensate of the
vector operator $\langle J_{x}\rangle=\rho_{x}^{(\Delta)}$ as a
function of temperature for the fixed masses of the vector field
$m^2r_{+}^{\theta}=0$ (left) and $m^2r_{+}^{\theta}=5/4$ (right). In
each panel, the three lines from bottom to top correspond to
increasing hyperscaling violation, i.e., $\theta=0$ (red), $1.0$
(black and dashed) and $1.9$ (blue) respectively. }
\end{figure}

In Fig. \ref{PWaveCond}, we present the condensate of the vector
operator $\langle J_{x}\rangle$ as a function of temperature for
different hyperscaling violation exponents $\theta$ with fixed
masses of the vector field $m^2r_{+}^{\theta}=0$ (left) and
$m^2r_{+}^{\theta}=5/4$ (right) in the holographic p-wave
superconductor model. Obviously, we find that the behavior of each
curve for the fixed $\theta$ and $m^2r_{+}^{\theta}$ agrees well
with the holographic superconducting phase transition in the
literature, which shows that the black hole solution with
non-trivial vector field can describe a superconducting phase.
Similar to the s-wave case in Fig. \ref{Condensate}, it is observed
that, for all cases considered here, the vector operator $\langle
J_{x}\rangle$ is single-valued near the critical temperature and the
condensate drops to zero continuously at the critical temperature.
Fitting these curves, we obtain a square root behavior for small
condensate
\begin{eqnarray}
\langle J_{x}\rangle\sim(1-T/T_{c})^{1/2}, \label{PWSquareRBehavior}
\end{eqnarray}
which means that the phase transition belongs to the second order
and the critical exponent of the system takes the mean-field value
$1/2$.

\begin{figure}[ht]
\includegraphics[scale=0.65]{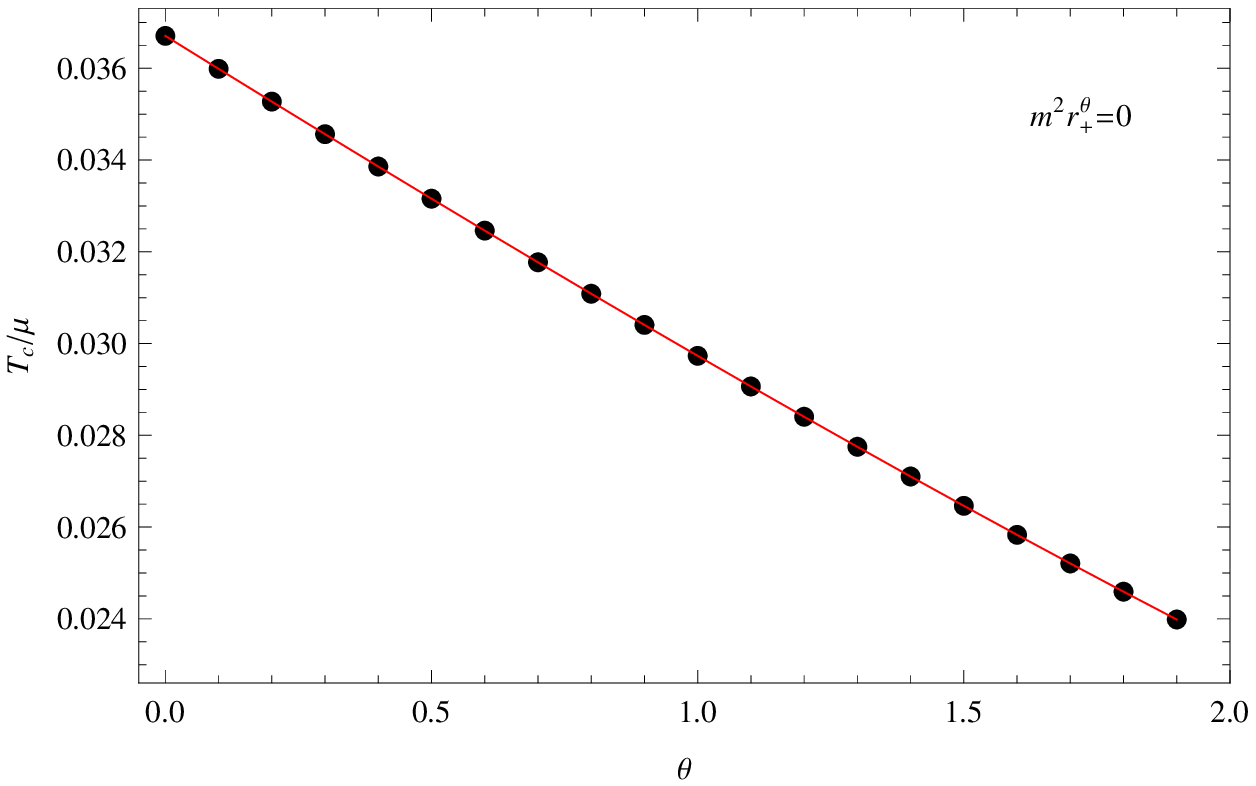}\hspace{0.2cm}%
\includegraphics[scale=0.65]{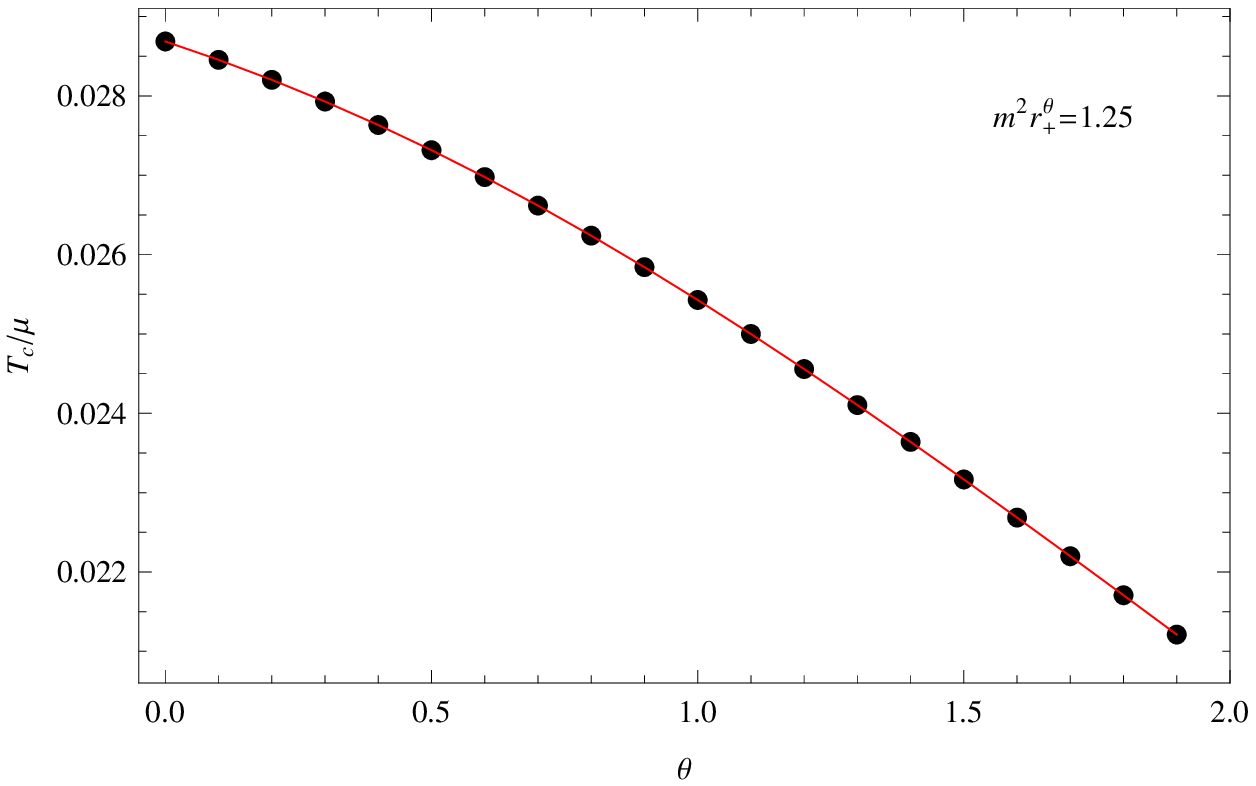}\\ \vspace{0.0cm}
\caption{\label{PWMaxwellTc} (Color online) The critical temperature
$T_{c}$ obtained by the numerical method as a function of the
hyperscaling violation $\theta$ with fixed masses of the vector
field $m^{2}r_{+}^{\theta}=0$ (left) and $m^{2}r_{+}^{\theta}=5/4$
(right) in the holographic p-wave superconductors with hyperscaling
violation.}
\end{figure}

In order to show the effect of the hyperscaling violation on the
critical temperature $T_{c}$ in the holographic p-wave
superconductors, in Fig. \ref{PWMaxwellTc} we plot the critical
temperature $T_{c}$ as a function of the hyperscaling violation
$\theta$ with fixed masses of the vector field
$m^{2}r_{+}^{\theta}=0$ (left) and $m^{2}r_{+}^{\theta}=5/4$
(right). We find that the critical temperature $T_{c}$ for the
vector operator $\langle J_{x}\rangle$ with the fixed vector field
mass decreases as the hyperscaling violation $\theta$ increases,
which implies that the higher hyperscaling violation makes it harder
for the condensation to form in the full parameter space
(\ref{Constraint}). Obviously, this behavior is different from that
seen for the holographic s-wave superconductors with hyperscaling
violation in Fig. \ref{CriticalTemperature}, where the critical
temperature $T_{c}$ decreases first and then increases as the
hyperscaling violation increases. Thus, we argue that, although the
underlying mechanism remains mysterious, the hyperscaling violation
has completely different effect on the critical temperature for the
s-wave and p-wave superconductor phase transitions.

\subsection{Conductivity}

In order to calculate the conductivity, we will consistently turn on
the perturbation of $A_{y}$ only and assume the form of the
perturbed Maxwell field $\delta A_{y}=A_{y}(u)e^{-i\omega t}dy$,
which leads to the equation of motion
\begin{eqnarray}
A_{y}^{\prime\prime}+\left(\frac{f^\prime}{f}-\frac{1}{u}\right)A_{y}^\prime
+\left(\frac{r_{+}^{4}\omega^2u^{2}}{f^2}-\frac{2r_{+}^{2}\rho_{x}^{2}}{f}\right)A_{y}=0.
\label{PWConductivityEquation}
\end{eqnarray}
Considering the ingoing wave boundary condition near the horizon
\begin{eqnarray}
A_{y}(u)\sim (1-u)^{-\frac{i\omega}{4\pi T}},
\end{eqnarray}
and the behavior in the asymptotic region ($u\rightarrow 0$)
\begin{eqnarray}
A_{y}=A_{y}^{(0)}+A_{y}^{(1)}r_{+}^{2}u^{2}.
\end{eqnarray}
we can express the conductivity as \cite{HartnollPRLJHEP}
\begin{eqnarray}\label{PWConductivity}
\sigma=-\frac{2iA_{y}^{(1)}}{\omega A_{y}^{(0)}}\ .
\end{eqnarray}
We still choose the masses of the vector field $m^2r_{+}^{\theta}=0$
and $m^2r_{+}^{\theta}=5/4$ in our calculation.

\begin{figure}[H]
\includegraphics[scale=0.425]{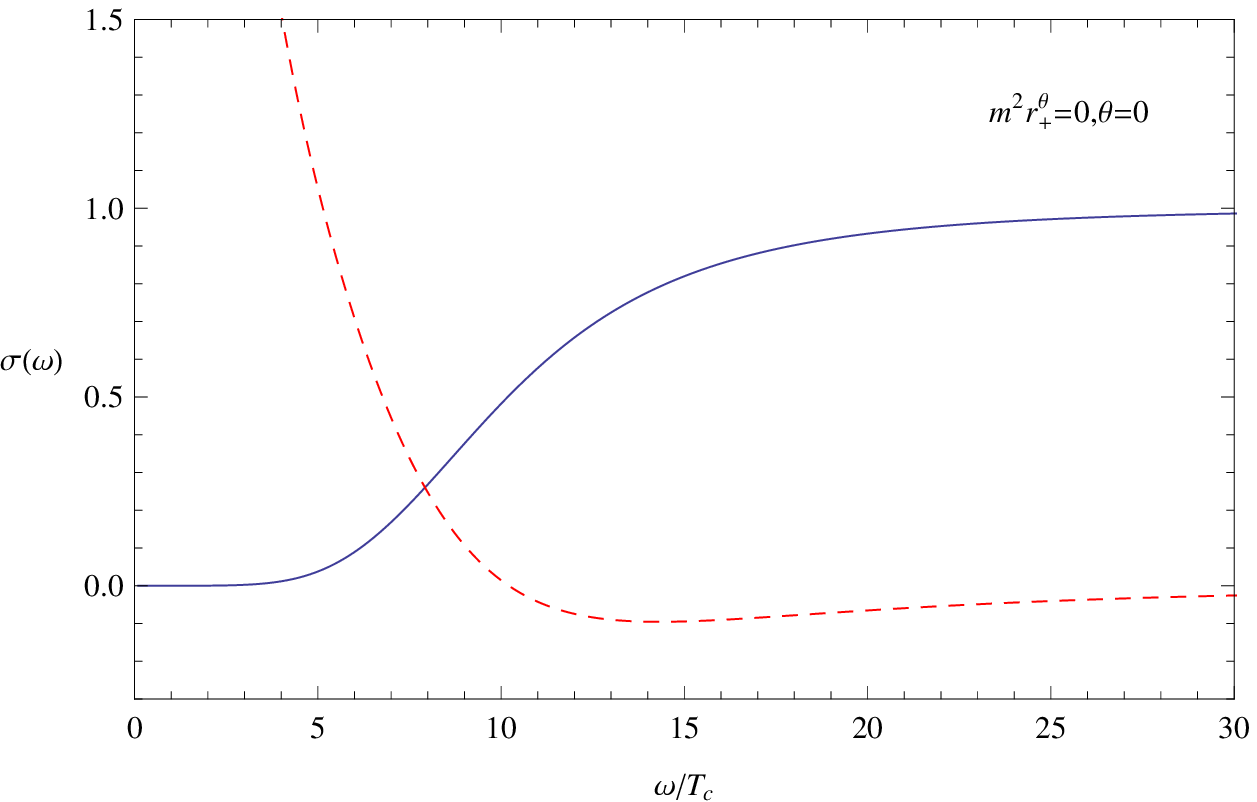}\hspace{0.2cm}%
\includegraphics[scale=0.425]{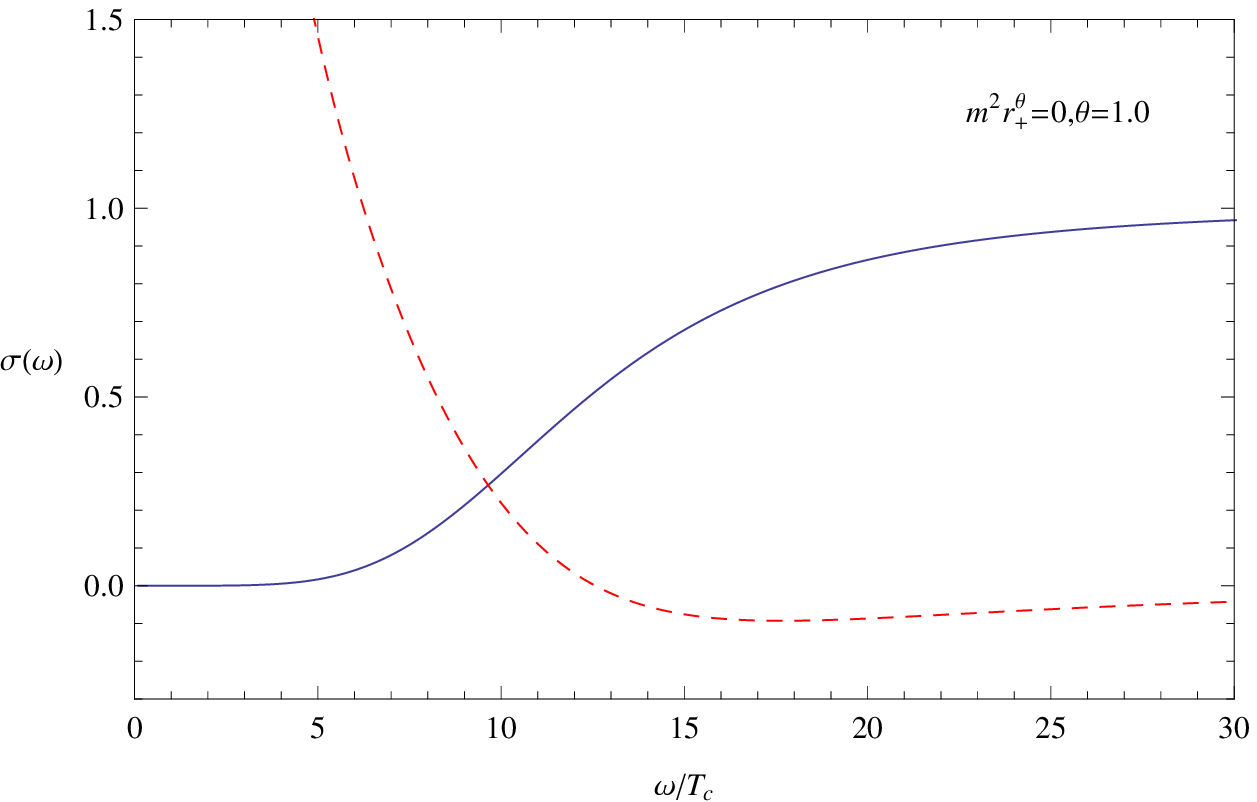}\hspace{0.2cm}%
\includegraphics[scale=0.425]{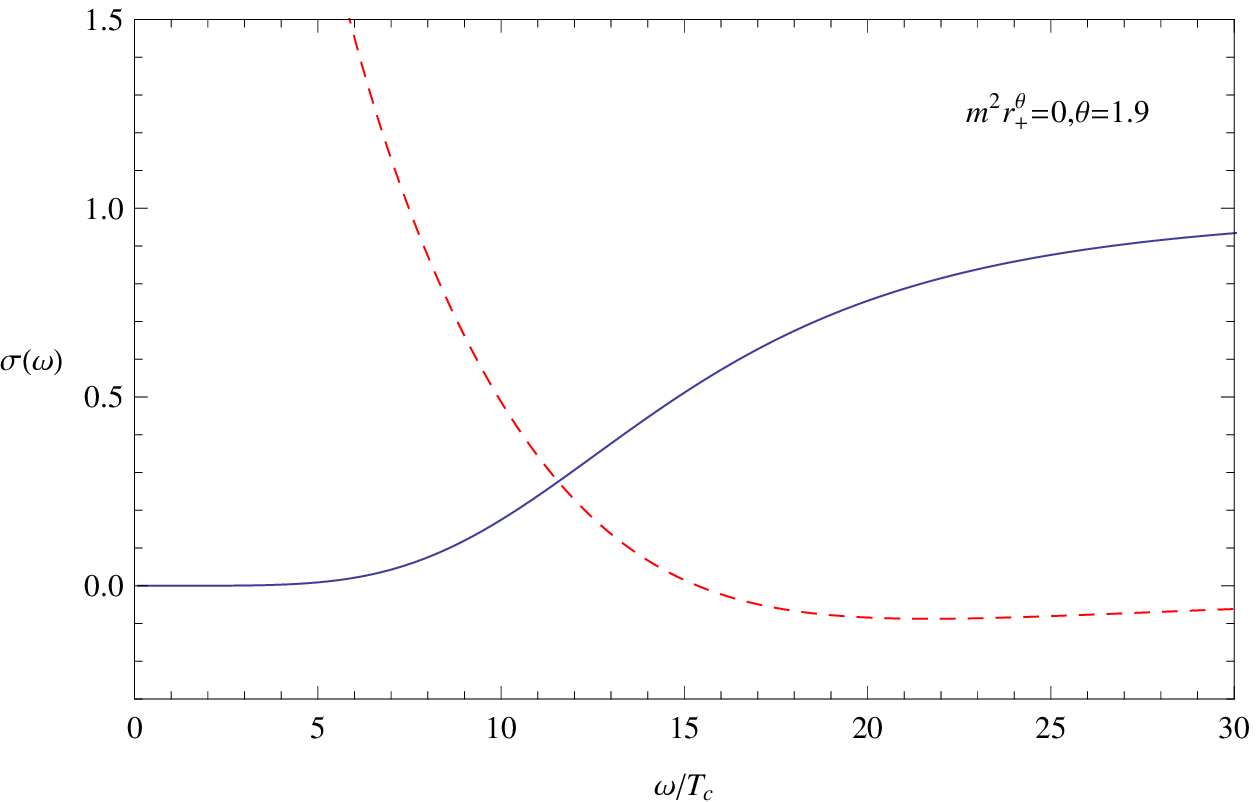}\\ \vspace{0.0cm}
\includegraphics[scale=0.425]{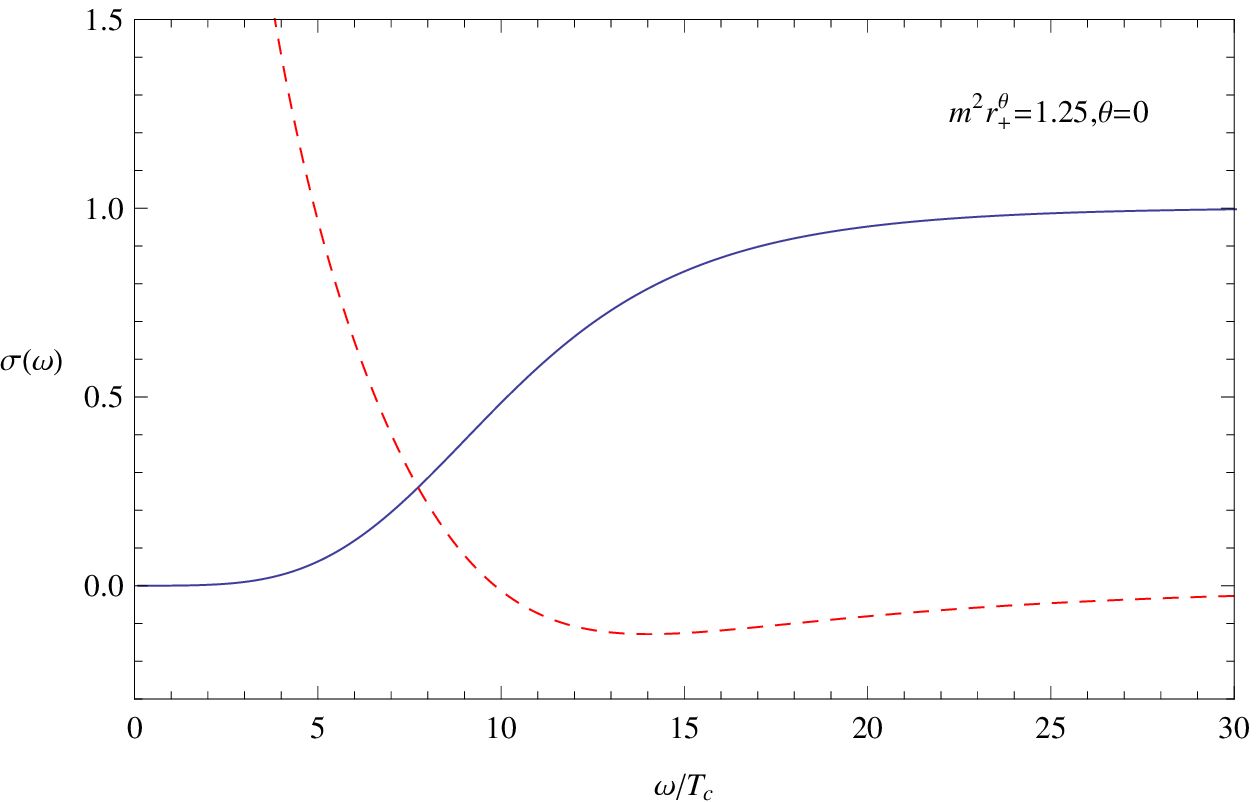}\hspace{0.2cm}%
\includegraphics[scale=0.425]{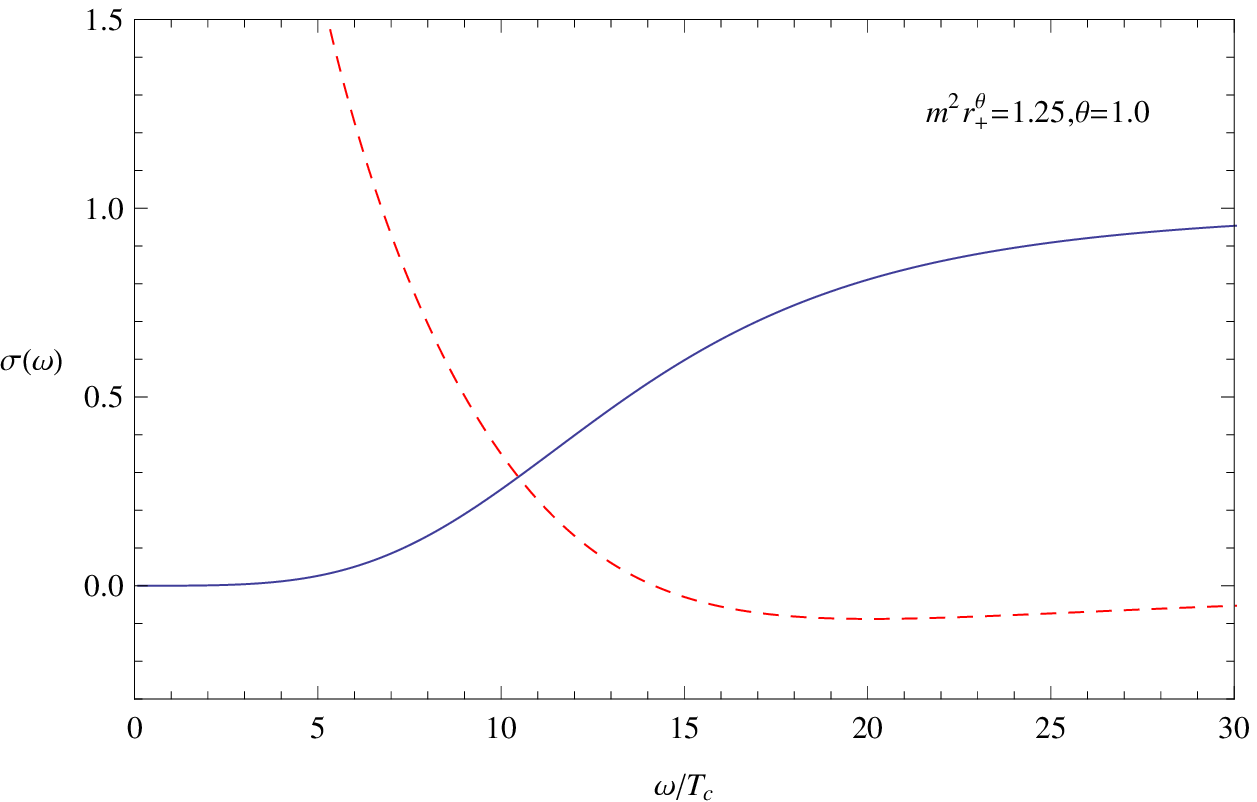}\hspace{0.2cm}%
\includegraphics[scale=0.425]{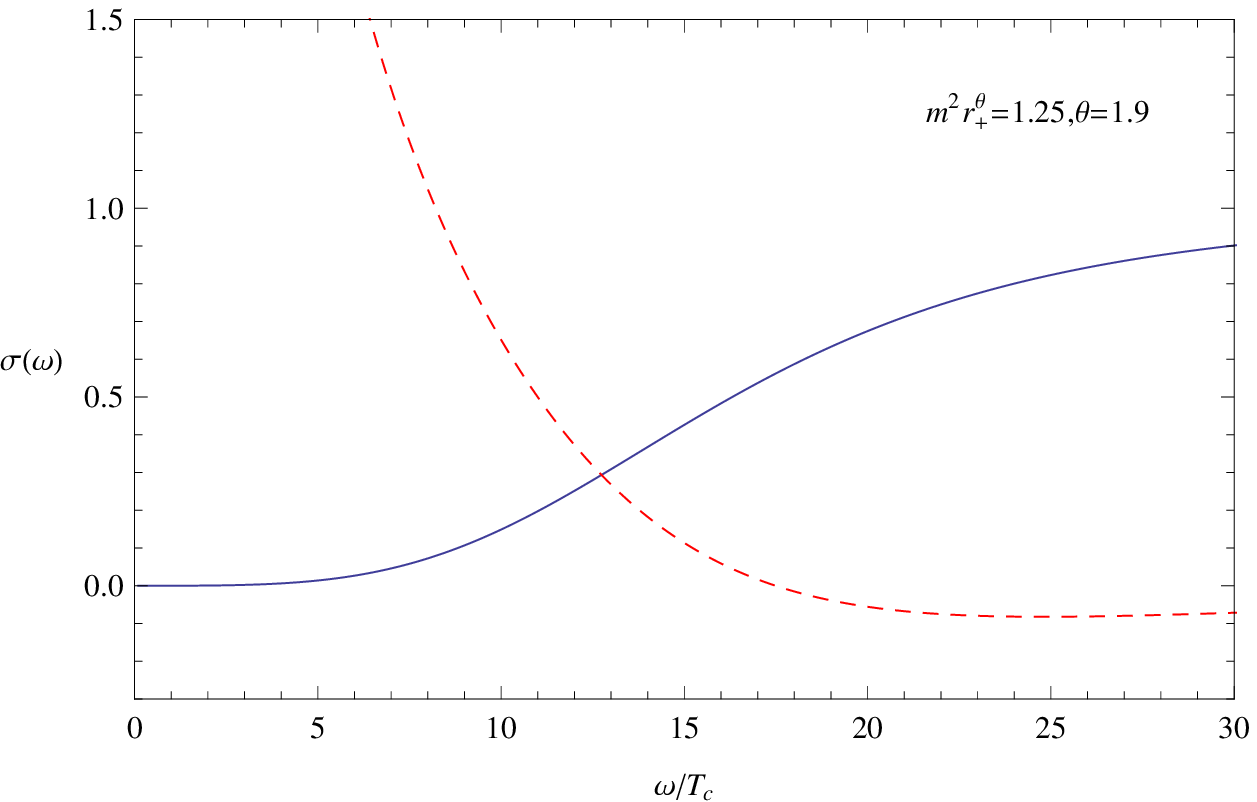}\\ \vspace{0.0cm}
\caption{\label{PWConductivity} (color online) Conductivity of the
holographic p-wave superconductors with hyperscaling violation for
the fixed masses of the vector field and different values of the
hyperscaling violation. In each panel, the blue (solid) line and red
(dashed) line represent the real part and imaginary part of the
conductivity $\sigma(\omega)$ respectively.}
\end{figure}

Solving the Maxwell equation (\ref{PWConductivityEquation})
numerically for $\theta=0$, $1.0$ and $1.9$ with
$m^2r_{+}^{\theta}=0$ and $m^2r_{+}^{\theta}=5/4$ at temperatures
$T/T_{c}\approx0.15$, we present the frequency dependent
conductivity of the holographic p-wave superconductors with
hyperscaling violation in Fig. \ref{PWConductivity}. Similar to the
s-wave case in Fig. \ref{Conductivity}, for all cases considered
here, we observe that the conductivity develops a gap with the gap
frequency $\omega_{g}$ and the larger deviation from the value
$\omega_g/T_c\approx 8$ as the hyperscaling violation $\theta$
increases. Therefore, we conclude that the higher hyperscaling
violation results in the larger deviation from the universal value
$\omega_g/T_c\approx 8$ \cite{HorowitzPRD78} for the gap frequency
in both s-wave and p-wave holographic superconductor models.

\section{Conclusions}

We have investigated the holographic superconductors with
hyperscaling violation in the probe limit, which may help to
understand the condensed matter materials with scaling properties
going beyond the standard Lorentz scaling at criticality. In the
s-wave (scalar field) model, different from the findings as shown in
Ref. \cite{FanJHEP} that the critical temperature increases as the
hyperscaling violation increases for the case of $z=2$ Lifshitz
scaling, we found that the critical temperature decreases first and
then increases with the increase of the hyperscaling violation,
which implies that the increase of the hyperscaling violation makes
the condensation of the scalar operator harder for small $\theta$
but easier for large $\theta$. Interestingly, the mass of the scalar
field will not modify the value of the hyperscaling violation
$\theta_{*}\approx0.6$ which gives the minimum critical temperature.
We improved the S-L method by including the higher order terms in
the expansion of the trial function to confirm our numerical results
and argued that, compared with the second order trial function used
in Ref. \cite{FanJHEP}, the higher order trial function can indeed
be used to give the analytical results which are completely
consistent with the numerical findings. However, the story is
completely different if we extend the investigation to the p-wave
(vector field) model. In contrast to the s-wave model, we observed
in the p-wave case that the critical temperature decreases as the
hyperscaling violation increases, which tells us that the higher
hyperscaling violation makes it harder for the vector condensation
to form in the full parameter space. Thus, we concluded that,
although the underlying mechanism remains mysterious, the
hyperscaling violation has completely different effect on the phase
transitions for the holographic s-wave and p-wave superconductors.
On the other hand, it should be noted that the Maxwell complex
vector model is still a generalization of the $SU(2)$ model even in
the hyperscaling violation geometry. Moreover, we pointed out that
the hyperscaling violation affects the conductivity of the
holographic superconductors and the higher hyperscaling violation
results in the larger deviation from the universal value
$\omega_g/T_c\approx 8$ for the gap frequency in both s-wave and
p-wave models. The extension of this work to the fully backreacted
spacetime would be interesting since the backreaction provides
richer physics in the holographic superconductor models. We will
leave it for further study.

\begin{acknowledgments}

We thank Professor Elcio Abdalla for his helpful discussions and
suggestions. This work was supported by FAPESP No. 2013/26173-9 and
CNPq (Brazil); the National Natural Science Foundation of China
under Grant No. 11275066; and the Open Project Program of State Key
Laboratory of Theoretical Physics, Institute of Theoretical Physics,
Chinese Academy of Sciences, China (No. Y5KF161CJ1).

\end{acknowledgments}

\appendix

\section{Equations of motion for $SU(2)$ p-wave superconductors with hyperscaling violation}

In this appendix, we will construct the holographic p-wave
superconductor models with hyperscaling violation by considering an
$SU(2)$ Yang-Mills action \cite{GubserPufu}
\begin{eqnarray}\label{pYM-System}
S=\int
d^{4}x\sqrt{-g}\left(-\frac{1}{4\hat{g}^2}F^{a}_{\mu\nu}F^{a\mu\nu}\right),
\end{eqnarray}
where $\hat{g}$ is the Yang-Mills coupling constant and
$F^{a}_{\mu\nu}=\partial_{\mu}A^{a}_{\nu}-\partial_{\nu}A^{a}_{\mu}+\epsilon^{abc}A^{b}_{\mu}A^{c}_{\nu}$
is the $SU(2)$ Yang-Mills field strength. The $A^{a}_{\mu}$ are the
components of the mixed-valued gauge fields
$A=A^{a}_{\mu}\tau^{a}dx^{\mu}$, where $\tau^{a}$ are the three
generators of the $SU(2)$ algebra with commutation relation
$[\tau^{a},\tau^{b}]=\epsilon^{abc}\tau^{c}$, and $\epsilon^{abc}$
is the totally antisymmetric tensor with $\epsilon^{123}=+1$.

We will take the ansatz of the gauge fields as \cite{GubserPufu}
\begin{eqnarray}\label{pYM-ansatz}
A(u)=\phi(u)\tau^{3}dt+\psi(u)\tau^{1}dx,
\end{eqnarray}
where we regard the $U(1)$ symmetry generated by $\tau^{3}$ as the
$U(1)$ subgroup of $SU(2)$ and the gauge boson with nonzero
component $\psi(u)$ along $x$-direction is charged under
$A^{3}_{t}=\phi(u)$. From the gauge/gravity duality, $\phi(u)$ is
dual to the chemical potential and $\psi(u)$ is dual to the
$x$-component of some charged vector operator $\langle J_{x}\rangle$
on the boundary. The condensation of $\psi(u)$ will spontaneously
break the $U(1)$ gauge symmetry and induce a phase transition, which
can be interpreted as a p-wave superconductor phase transition in
the boundary field theory.

From the Yang-Mills action (\ref{pYM-System}) and the solution
(\ref{SolitonU}) for $d=2,~z=2$, we have the following equations of
motion
\begin{eqnarray}
\psi^{\prime\prime}
+\left(\frac{f^\prime}{f}-\frac{1}{u}\right)\psi^\prime
+\frac{r_{+}^{4}u^{2}\phi^2}{f^2}\psi=0, \label{pYMwave-Psi}
\end{eqnarray}
\begin{eqnarray}
\phi^{\prime\prime}+\frac{1}{u}\phi^\prime-\frac{r_{+}^{2}\psi^2}{f}\phi=0,
\label{pYMwave-Phi}
\end{eqnarray}
where the prime denotes the derivative with respect to $u$.

\end{document}